\documentclass[11pt,a4paper]{article}
\pdfoutput=1

\usepackage{jheppub}
\usepackage[T1]{fontenc}
\usepackage{lmodern}
\usepackage{mathtools}
\usepackage{booktabs}
\usepackage{microtype}
\usepackage{xcolor}
\usepackage{braket}
\usepackage{float}
\usepackage{placeins}

\newcommand{\DG}{\mathrm{DG}}
\newcommand{\TW}{\mathrm{TW}}
\newcommand{\Prob}{\mathbb P}
\newcommand{\dd}{\mathrm d}
\newcommand{\ee}{\mathrm e}
\newcommand{\tr}{\operatorname{tr}}
\newcommand{\Var}{\operatorname{Var}}
\newcommand{\cTW}{C_{\mathrm{TW}}}
\newcommand{\TWmean}{\langle x\rangle_{\mathrm{TW}_2}}
\newcommand{\Unif}{\operatorname{Unif}}
\newcommand{\Phihat}{\widehat\Phi}

\newcommand{\nn}{\nonumber}

\allowdisplaybreaks
\title{The Hagedorn transition as a Plancherel wall: Tracy--Widom statistics, Polyakov loops and partial deconfinement}

\author[a,b]{Robert de Mello Koch,}
\author[a,c]{Minkyoo Kim}
\author[d]{and Hyunwoo Oh}

\affiliation[a]{School of Science, Huzhou Normal University, Huzhou 313000, China}
\affiliation[b]{Mandelstam Institute for Theoretical Physics, School of Physics,\\
University of the Witwatersrand, Private Bag 3, Wits 2050, South Africa}
\affiliation[c]{Research Institute of Basic Sciences, Seoul National University, Seoul 08826, Korea}
\affiliation[d]{Department of Physics and Astronomy \& Center for Theoretical Physics,\\
Seoul National University, Seoul 08826, Korea}

\emailAdd{robert@zjhu.edu.cn}
\emailAdd{mimkim80@gmail.com}
\emailAdd{hyunwoo1535@snu.ac.kr}

\abstract{We study the Hagedorn transition of the single-winding Dutta--Gopakumar unitary matrix model at finite rank $N$. Its Schur expansion is a sum of Plancherel probabilities restricted by the finite-rank wall on the number of rows, and the Vershik--Kerov--Logan--Shepp limit shape reaches this wall when the number of boxes equals $N^2/4$. Two scaling windows around the Hagedorn point resolve the wall at different scales. In the outer window the Baik--Deift--Johansson theorem gives a Tracy--Widom crossover, and summing it identifies Liu's critical free-energy coefficient with the first moment of the GUE Tracy--Widom distribution. In the inner window the same sum determines the canonical probability law of the Polyakov loop at the Hagedorn point: the normalized loop $\tr U /N$ is uniformly distributed on a disk of radius one half, so the order parameter does not self-average at large $N$. 
The intensity $|\tr U|^2/N^2$ has the same limiting law as the Schur degree $n/N^2$: at the Hagedorn point the canonical degree weights become asymptotically equal below the macroscopic finite-rank wall. Under the partial-deconfinement dictionary the uniform disk becomes a linear probability density for the deconfined color fraction, and its fully deconfined endpoint is the Young-diagram endpoint of Berenstein and Yan. The Tracy–Widom mean also fixes the first finite-\(N\) correction to the Polyakov-loop Laplace transform and to all of its fixed radial moments. Bessel–Toeplitz evaluations up to $N = 100$ confirm the free-energy coefficient, the uniform law, and the predicted correction.}

\begin{document}
\maketitle
\flushbottom

\section{Introduction}
\label{sec:intro}

The Hagedorn transition of free $\mathcal N=4$ super Yang--Mills theory on
$S^3$ is a sharp large-$N$ phenomenon: the planar partition function diverges
when the first-winding coefficient, given by the single-particle partition
function, reaches unity \cite{Sundborg1999,Aharony2004}. At finite rank the divergence is smoothed. In
the single-winding truncation studied by Dutta and Gopakumar
\cite{DuttaGopakumar2008}, the relevant unitary matrix integral is
\begin{equation}
 Z_{\DG,N}(a)=\int_{U(N)}[\dd U]\,\exp\!\left(a|\tr U|^2\right),
 \label{eq:DG-intro}
\end{equation}
where $a=1$ is the Hagedorn point. The planar confined answer
$Z_{\DG,\infty}(a)=(1-a)^{-1}$ has a pole there, whereas the finite-$N$
integral is regular. Liu described its smoothing in a double-scaling limit in
terms of a Painlev\'e-II function $B_0(s)$ \cite{Liu2005}.

The Schur expansion gives a direct finite-rank interpretation. At fixed degree
$n$, the coefficient of $a^n$ is the Plancherel probability that a Young
diagram has at most $N$ rows,
\begin{equation}
 Z_{\DG,N}(a)=\sum_{n\geq0}a^n p_N(n),
 \qquad p_N(n)=\Prob_n(\lambda'_1\leq N).
 \label{eq:intro-schur}
\end{equation}
The restriction $\lambda'_1\leq N$ is the character-basis form of the
finite-$N$ trace relations. In the single-winding model, all rank dependence of the exact Schur sum
is carried by this restriction; its interpretation in terms of trace
relations is consistent with the finite-$N$ invariant description
\cite{deMelloKochJevicki2025a,CorleyJevickiRamgoolam2002,Procesi1976}. It also relates the two saddle families of
\cite{DuttaGopakumar2008}: the unconstrained Vershik--Kerov--Logan--Shepp (VKLS) Young diagram continues into
a wall-saturated, truncated VKLS (tVKLS) Young diagram once its depth reaches
$N$ \cite{VershikKerov1977,LoganShepp1977,DeMelloKochKimLarweh2024}.

The first result concerns the outer scaling window $1-a\sim N^{-4/3}$. The VKLS edge $\lambda'_1\simeq2\sqrt n$ reaches the wall at $n=N^2/4$,
and the Baik--Deift--Johansson (BDJ) theorem fixes the local crossover
\cite{BaikDeiftJohansson1999,TracyWidom1994,BorodinOkounkovOlshanski2000,Johansson2001}.
Summing this crossover gives
\begin{equation}
 \log Z_{\DG,N}(1)
 =2\log N-\log4+2^{2/3}|\TWmean|N^{-2/3}+O(N^{-4/3}),
 \qquad
 4B_0(0)=2^{2/3}|\TWmean|,
 \label{eq:intro-free}
\end{equation}
where $\TWmean=-1.7710868074\ldots$ in the standard GUE normalization
\cite{Bornemann2010}. Thus Liu's amplitude is a soft-edge moment rather than
an independent matrix-model constant.

The second result comes from the narrower inner window $1-a\sim N^{-2}$.
Writing $a=1-t/N^2$, the ratio
$Z_{\DG,N}(1-t/N^2)/Z_{\DG,N}(1)$ is exactly the Laplace transform of
$Q_N=|\tr U|^2/N^2$ in the critical canonical ensemble. The sharp macroscopic
Plancherel wall then gives
\begin{equation}
 Q_N\longrightarrow\Unif\!\left[0,\tfrac14\right],
 \qquad
 \frac{\tr U}{N}\ \hbox{is uniform on } |P|\leq\frac12.
 \label{eq:intro-uniform}
\end{equation}
The distribution has an $O(1)$ width at $N=\infty$, so the order parameter is
not self-averaging. The same edge argument predicts the first correction to
the fixed-$t$ transform and to every fixed radial moment,
\begin{equation}
 \langle |P|^{2m}\rangle_{a=1}
 =\frac{1}{(m+1)4^m}
 +\frac{m\cTW}{(m+1)4^{m-1}}N^{-2/3}+o(N^{-2/3}),
 \qquad
 \cTW=\frac{|\TWmean|}{2^{4/3}}=B_0(0).
 \label{eq:intro-moments}
\end{equation}
We test this prediction directly with the Bessel--Toeplitz representation.

The critical disk has a partial-deconfinement interpretation. The standard eigenvalue density of a deconfined $U(M)$ block and a confined
remainder is the gapless flat family of the critical matrix model
\cite{Aharony2004,HanadaIshikiWatanabe2019,Berenstein2018,HanadaWatanabe2023}. With
$A=M/N$, it gives $|P|=A/2$. Therefore the uniform law for $Q=|P|^2$ induces
\begin{equation}
 \rho_A(A)=2A,\qquad 0\leq A\leq1.
 \label{eq:intro-partial}
\end{equation}
The canonical ensemble samples the full continuum of partially deconfined
sectors, uniformly in $A^2$. Its fully deconfined endpoint $A=1$ is the same
finite-rank wall that carries the Tracy--Widom crossover; this endpoint has
also been emphasized in related free matrix models \cite{BerensteinYan2023}.

The paper is organized as follows. Section~\ref{sec:wall} derives the exact
Schur decomposition and identifies the VKLS/tVKLS crossover.
Section~\ref{sec:free-energy} obtains the Hagedorn amplitude and relates it to
$B_0(s)$. Section~\ref{sec:polyakov} derives the critical Polyakov-loop law,
its partial-deconfinement interpretation, and the Tracy--Widom correction at
the fully deconfined endpoint. Section~\ref{sec:numerics} presents the
finite-$N$ checks, and Section~\ref{sec:discussion} discusses the scope and
extensions of the result. The appendices give the probability-measure
formulation, the wall-control argument, and the normalization conventions.

\section{The Schur wall}
\label{sec:wall}

\subsection{Exact decomposition and the Plancherel wall}
\label{sec:schur-decomposition}

For a Young diagram $R\vdash n$, let $d_R$ be the dimension of the
corresponding $S_n$ irrep and let $\chi_R(U)$ be the $U(N)$ character. The
Frobenius character formula and Schur orthogonality are \cite{CorleyJevickiRamgoolam2002}
\begin{equation}
 (\tr U)^n=\sum_{R\vdash n}d_R\chi_R(U),
 \qquad
 \int_{U(N)}[\dd U]\,\chi_R(U)\chi_S(U^\dagger)
 =\delta_{RS}[\ell(R)\leq N].
 \label{eq:frobenius-orthogonality}
\end{equation}
Expanding \eqref{eq:DG-intro} therefore gives the exact identity
\begin{equation}
 \begin{split}
 Z_{\DG,N}(a)
 &=\sum_{n=0}^{\infty}\frac{a^n}{n!}
   \int[\dd U]\,|\tr U|^{2n}\\
 &=\sum_{n=0}^{\infty}a^n
   \sum_{\substack{R\vdash n\\ \ell(R)\leq N}}\frac{d_R^2}{n!}
 =\sum_{n=0}^{\infty}a^n p_N(n),
 \end{split}
 \label{eq:exact-schur}
\end{equation}
where
\begin{equation}
 p_N(n)=\Prob_n(\lambda'_1\leq N),
 \qquad
 \Prob_n(R)=\frac{d_R^2}{n!}
 \label{eq:plancherel-probability}
\end{equation}
is Plancherel measure on diagrams with $n$ boxes. Removing the wall replaces
$p_N(n)$ by one and recovers the planar geometric series. For fixed $N$ the
series is still absolutely convergent at $a=1$, since
\begin{equation}
 p_N(n)=\frac1{n!}\int[\dd U]\,|\tr U|^{2n}
 \leq\frac{N^{2n}}{n!}.
 \label{eq:absolute-bound}
\end{equation}

The wall is not an auxiliary cutoff. The Schur polynomial vanishes when $\ell(R)>N$, so the row wall implements
the finite-rank restriction in the character basis
\cite{CorleyJevickiRamgoolam2002}.
For the DG integral, no other $N$-dependent factor occurs in the exact
coefficient \eqref{eq:plancherel-probability}.
For a thermal model with several adjoint letters, the same row restriction
remains, while representation multiplicities modify the weights
\cite{BhattacharyyaCollinsDeMelloKoch2008,BrownHeslopRamgoolam2008};
we return to the distinction between letter content and winding truncation
in Section~\ref{sec:discussion}.

\subsection{VKLS, truncated VKLS, and the saddle identification}
\label{sec:vkls-saddles}

Plancherel measure is invariant under transposition, so the first-column length
$\lambda'_1$ has the same law as the longest row. The VKLS limit shape gives \cite{VershikKerov1977,LoganShepp1977,BorodinOkounkovOlshanski2000}
\begin{equation}
 \frac{\lambda'_1}{\sqrt n}\longrightarrow2
 \qquad(n\longrightarrow\infty).
 \label{eq:vkls-edge}
\end{equation}
The unconstrained Young diagram reaches the finite-rank wall at
\begin{equation}
 n_*=\frac{N^2}{4}.
 \label{eq:nstar}
\end{equation}
Below $n_*$ the typical diagram is the ordinary VKLS diagram. Above $n_*$,
diagrams contributing to \eqref{eq:exact-schur} must press against the row wall
and spread horizontally. This tVKLS continuation is the representation-space
counterpart of the second large-$N$ saddle family of the unitary matrix model.

Shimada and Watanabe derive the VKLS profile in the partially deconfined
regime of the Gaussian multi-matrix model without truncating the
higher-winding couplings, and identify the number of rows of the dominant
Young diagram with the size of the deconfined subsector
\cite{Shimada:2026ipv}.
Their analysis determines the profile at fixed scale, while the canonical
measure over this family in the single-winding model is determined below.

The crossover width follows directly. A displacement $\delta n$ moves the
classical edge by $\delta(2\sqrt n)\simeq 2\delta n/N$, while the Plancherel edge
fluctuates on the $n^{1/6}\sim N^{1/3}$ scale. Hence
$\delta n\sim N^{4/3}$, which is also the origin of Liu's scaling
$1-a\sim N^{-4/3}$.

\section{The free-energy amplitude}
\label{sec:free-energy}

\subsection{Tracy--Widom resummation at the Hagedorn point}
\label{sec:tw-resummation}

Let $F_2(x)$ be the GUE Tracy--Widom distribution function. The BDJ theorem
states that, for fixed $x$,
\begin{equation}
 \Prob_n\!\left(\lambda'_1\leq2\sqrt n+xn^{1/6}\right)
 \longrightarrow F_2(x).
 \label{eq:bdj}
\end{equation}
In the crossover region write
\begin{equation}
 n=\frac{N^2}{4}+N^{4/3}\tau.
 \label{eq:tau-def}
\end{equation}
Then
\begin{equation}
 \frac{N-2\sqrt n}{n^{1/6}}
 =-2^{4/3}\tau+O(N^{-2/3}),
 \label{eq:wall-coordinate}
\end{equation}
and therefore
\begin{equation}
 p_N\!\left(\frac{N^2}{4}+N^{4/3}\tau\right)
 \longrightarrow F_2(-2^{4/3}\tau).
 \label{eq:wall-profile}
\end{equation}

Subtracting the sharp VKLS step gives
\begin{equation}
 Z_{\DG,N}(1)
 =\sum_{n\geq0}\theta(n_*-n)
 +\sum_{n\geq0}\bigl[p_N(n)-\theta(n_*-n)\bigr].
 \label{eq:step-subtraction}
\end{equation}
The first term is $N^2/4+O(1)$. At leading order the second becomes a
Riemann sum across the $N^{4/3}$ edge window,
\begin{equation}
 \sum_{n\geq0}\bigl[p_N(n)-\theta(n_*-n)\bigr]
 \sim N^{4/3}\mathcal I_{\TW},
 \qquad
 \mathcal I_{\TW}=\int_{-\infty}^{\infty}
 \bigl[F_2(-2^{4/3}\tau)-\theta(-\tau)\bigr]\dd\tau.
 \label{eq:edge-riemann-sum}
\end{equation}
Changing variables and using
$\int_{-\infty}^{\infty}[F(x)-\theta(x)]\dd x=-\langle x\rangle_F$ gives
\begin{equation}
 \mathcal I_{\TW}=-\frac{\TWmean}{2^{4/3}}
 =\frac{|\TWmean|}{2^{4/3}}.
 \label{eq:tw-edge-integral}
\end{equation}
Thus
\begin{align}
 Z_{\DG,N}(1)&=\frac{N^2}{4}+\cTW N^{4/3}+O(N^{2/3}),
 \label{eq:res1-Z}\\
 \log Z_{\DG,N}(1)&=2\log N-\log4+4\cTW N^{-2/3}+O(N^{-4/3}),
 \label{eq:res1-log}
\end{align}
with
\begin{equation}
 \cTW=\frac{|\TWmean|}{2^{4/3}}=0.702856\ldots,
 \qquad
 4\cTW=2^{2/3}|\TWmean|=2.811425\ldots.
 \label{eq:ctw-value}
\end{equation}

\subsection{Relation to Liu's scaling function}
\label{sec:liu}

Liu's double-scaling variable is
\begin{equation}
 1-a=\frac{s}{N^{4/3}},
 \label{eq:liu-variable}
\end{equation}
with $s$ fixed as $N\longrightarrow\infty$. In this convention
\cite{Liu2005},
\begin{equation}
 Z_{\DG,N}(a)
 =\frac{N^{4/3}}{s}\left(1-\ee^{-sN^{2/3}/4}\right)
 +N^{4/3}\ee^{-sN^{2/3}/4}B_0(s)+O(N^{2/3}).
 \label{eq:liu-expansion}
\end{equation}
At $s=0$ the first term tends to $N^2/4$, and comparison with
\eqref{eq:res1-Z} gives $B_0(0)=\cTW$.

Liu's Painlev\'e-II representation can be written as
\begin{equation}
 B_0(s)=\frac12\int_{-\infty}^{\infty}\dd t\,
 \ee^{st/2}\left[\ee^{F^{(2)}_0(t)}-\theta(t)\right],
 \label{eq:liu-B0}
\end{equation}
where $\ee^{F^{(2)}_0(t)}=F_2(2^{1/3}t)$. Hence
\begin{equation}
 B_0(s)=\frac{1}{2^{4/3}}
 \int_{-\infty}^{\infty}\dd x\,
 \ee^{sx/2^{4/3}}\left[F_2(x)-\theta(x)\right].
 \label{eq:B0-transform}
\end{equation}
Thus $B_0(s)$ is the bilateral Laplace transform of the Tracy--Widom CDF
excess over the sharp planar step. In particular,
\begin{equation}
 B_0(0)=\frac{|\TWmean|}{2^{4/3}}=\cTW,
 \qquad
 4B_0(0)=2^{2/3}|\TWmean|.
 \label{eq:B0-zero}
\end{equation}

\section{Critical Polyakov-loop law and partial deconfinement}
\label{sec:polyakov}

We want to extract the large-$N$ probability distribution of the Polyakov
loop in the Dutta--Gopakumar model at the Hagedorn point. In the canonical
ensemble at $a=1$,
\begin{equation}
 \frac{|\tr U|^2}{N^2}\xrightarrow{\ d\ }\Unif\!\left[0,\tfrac14\right].
 \label{eq:sec4-statement}
\end{equation}
This result follows from the exact Schur--Plancherel representation together
with the sharp large-$N$ Plancherel wall at $n=N^2/4$. It reveals a new piece
of critical physics: the order parameter is non-self-averaging at large $N$,
and the critical ensemble samples a continuum of partially deconfined
sectors. The Tracy--Widom mean then controls the leading finite-$N$
correction to the full order-parameter distribution, rather than only the
correction to the free energy.

In Section~\ref{sec:free-energy} we identified
 the leading finite-$N$ smoothing of the Hagedorn transition with
the Tracy--Widom mean; it derives the critical behavior of the partition
function and writes Liu's scaling function as a Laplace transform of the
excess of the Tracy--Widom cumulative distribution over the sharp planar step
profile. The same model contains additional information about the
distribution of the deconfinement order parameter. The relevant scaling regime
is not the Tracy--Widom ``outer window'' ($1-a\sim N^{-4/3}$), but the
narrower ``inner window'' $1-a\sim N^{-2}$. This inner scaling probes
$|\tr U|^2/N^2$ in the canonical ensemble at the Hagedorn point.

\subsection{The inner window and the uniform law}
\label{sec:inner-window}

\paragraph{The critical Dutta--Gopakumar ensemble.}
At $a=1$, the normalized probability measure is
\begin{equation}
 \dd\mu_N(U)=\frac{1}{Z_{\DG,N}(1)}[\dd U]\,\ee^{|\tr U|^2}.
 \label{eq:critical-measure}
\end{equation}
The normalized Polyakov-loop intensity is
\begin{equation}
 Q_N=\frac{|\tr U|^2}{N^2}.
 \label{eq:PQ-def}
\end{equation}
Its Laplace transform in the critical ensemble is
\begin{equation}
 \Phihat_N(t)=\left\langle\ee^{-tQ_N}\right\rangle_{a=1}
 =\frac{1}{Z_{\DG,N}(1)}\int[\dd U]\,
 \exp\!\left[\left(1-\frac{t}{N^2}\right)|\tr U|^2\right]
 =\frac{Z_{\DG,N}\!\left(1-\frac{t}{N^2}\right)}{Z_{\DG,N}(1)}.
 \label{eq:exact-laplace}
\end{equation}
In the same way, for a fixed positive integer $m$,
\begin{equation}
 \langle Q_N^m\rangle_{a=1}
 =\frac{1}{N^{2m}}
 \frac{\sum_{n\geq0}(n)_m\,p_N(n)}{\sum_{n\geq0}p_N(n)},
 \qquad (n)_m=n(n-1)\cdots(n-m+1).
 \label{eq:exact-moment}
\end{equation}
Thus the complete critical order-parameter distribution is encoded in the
partition function evaluated in the inner scaling regime $a=1-t/N^2$.

\paragraph{The Plancherel-wall derivation.}
The exact Schur decomposition \eqref{eq:exact-schur} is
\begin{equation}
 Z_{\DG,N}(a)=\sum_{n\geq0}a^n p_N(n),
 \qquad
 p_N(n)=\Prob_n(\lambda'_1\leq N),
 \nonumber
\end{equation}
where $\Prob_n$ is Plancherel measure on Young diagrams with $n$ boxes and
$\lambda'_1$ is the first-column length. At large $N$, the Plancherel edge
reaches the finite-rank wall at $n_*=N^2/4$. On the macroscopic $n\sim N^2$
scale, the wall probability becomes a sharp step:
\begin{equation}
 p_N(N^2x)\longrightarrow
 \begin{cases}
 1,&0\leq x<\tfrac14,\\[2pt]
 0,&x>\tfrac14.
 \end{cases}
 \label{eq:sharp-step}
\end{equation}
The transition region has width $N^{4/3}$, which is negligible compared with
the $N^2$ scale relevant here. Set
\begin{equation}
 a_N=1-\frac{t}{N^2},
 \label{eq:aN}
\end{equation}
with fixed $t\geq0$. Equations \eqref{eq:intro-schur} and
\eqref{eq:sharp-step} imply
\begin{equation}
 Z_{\DG,N}(a_N)\simeq\sum_{n=0}^{N^2/4}\left(1-\frac{t}{N^2}\right)^n.
 \label{eq:truncated-geometric}
\end{equation}
Writing $n=N^2x$, in the $N\to\infty$ limit we find
\begin{equation}
 \left(1-\frac{t}{N^2}\right)^n\longrightarrow\ee^{-tx}.
 \label{eq:exp-weight}
\end{equation}
Consequently, the sum becomes an integral
\begin{equation}
 \frac{1}{N^2}Z_{\DG,N}\!\left(1-\frac{t}{N^2}\right)
 \longrightarrow\int_0^{1/4}\ee^{-tx}\dd x
 =\frac{1-\ee^{-t/4}}{t}.
 \label{eq:sharp-integral}
\end{equation}
At $t=0$, this reduces continuously to
\begin{equation}
 Z_{\DG,N}(1)\sim\frac{N^2}{4}.
 \label{eq:Z1-leading}
\end{equation}
Substituting into \eqref{eq:exact-laplace} gives (in the $N\to\infty$ limit)
\begin{equation}
 \Phihat_N(t)\longrightarrow\frac{4(1-\ee^{-t/4})}{t}.
 \label{eq:uniform-laplace}
\end{equation}
Now, notice that
\begin{equation}
 \frac{4(1-\ee^{-t/4})}{t}=4\int_0^{1/4}\ee^{-tq}\dd q
 \label{eq:uniform-laplace-integral}
\end{equation}
is the Laplace transform of the uniform probability
density\footnote{$\mathbf 1_{[0,1/4]}(q)$ is the indicator function of the
interval $[0,\tfrac14]$. It is $1$ if $0\leq q\leq\tfrac14$ and zero
otherwise.}
\begin{equation}
 \rho_Q(q)=4\,\mathbf 1_{[0,1/4]}(q).
 \label{eq:uniform-density}
\end{equation}
We therefore arrive at our limiting law
\begin{equation}
 Q_N=\frac{|\tr U|^2}{N^2}\xrightarrow{\ d\ }Q,
 \qquad
 Q\sim\Unif\!\left[0,\tfrac14\right].
 \label{eq:uniform-law}
\end{equation}
Appendix~\ref{app:weak-concentration} proves \eqref{eq:uniform-law} using
VKLS concentration, monotonicity of $p_N(n)$, and the elementary far-tail
bound \eqref{eq:absolute-bound}.
The stronger large-deviation estimates discussed in
Appendix~\ref{app:uniform-large-deviation} are therefore not needed for the
leading uniform law.

\paragraph{Critical non-self-averaging.}
The limiting moments are
\begin{equation}
 \langle Q^k\rangle=4\int_0^{1/4}q^k\dd q=\frac{1}{4^k(k+1)}
 \quad\Rightarrow\quad
 \langle Q\rangle=\frac18,
 \qquad
 \langle Q^2\rangle=\frac1{48},
 \label{eq:uniform-moments}
\end{equation}
and hence
\begin{equation}
 \Var(Q)=\frac1{48}-\frac1{64}=\frac1{192}
 \quad\Rightarrow\quad
 \frac{\Var(Q)}{\langle Q\rangle^2}=\frac13.
 \label{eq:nonself-ratio}
\end{equation}
It remains finite as $N\to\infty$. Thus the Polyakov-loop intensity is not
self-averaging at the Hagedorn point. This is stronger than the statement
that two competing saddles coexist. The canonical ensemble does not
concentrate on a single value of the order parameter, or even on two isolated
values. It samples the entire continuum
\begin{equation}
 0\leq\frac{|\tr U|^2}{N^2}\leq\frac14.
 \label{eq:continuum}
\end{equation}
This broad distribution is a natural canonical signature of the flat Hagedorn
direction.

\subsection{Partial deconfinement and critical fluctuations}
\label{sec:partial-deconfinement}

\paragraph{Interpretation in terms of partial deconfinement.}
In the partial-deconfinement picture, one introduces a continuously varying
deconfined color fraction
\begin{equation}
 A=\frac{M}{N},\qquad0\leq A\leq1,
 \label{eq:A-def}
\end{equation}
where an $SU(M)$ or $U(M)$ subsector is deconfined while the remaining colors
stay confined \cite{HanadaIshikiWatanabe2019,HanadaMaltz2017,Berenstein2018,HanadaJevickiPengWintergerst2019,HanadaWatanabe2023}. The standard eigenvalue-density
ansatz is
\begin{equation}
 \rho(\theta)=(1-A)\rho_{\rm conf}(\theta)+A\rho_{\rm GWW}(\theta),
 \label{eq:pd-ansatz}
\end{equation}
with
\begin{equation}
 \rho_{\rm conf}(\theta)=\frac1{2\pi},
 \qquad
 \rho_{\rm GWW}(\theta)=\frac{1+\cos\theta}{2\pi}.
 \label{eq:pd-densities}
\end{equation}
Choosing the center phase so that $P$ is real and nonnegative, this ansatz gives
\begin{equation}
 P=\frac{\tr U}{N}=\int_{-\pi}^{\pi}\dd\theta\,\rho(\theta)\,\ee^{i\theta}
 =\frac{A}{2}.
 \label{eq:pd-dictionary}
\end{equation}
Therefore,
\begin{equation}
 Q=|P|^2=\frac{A^2}{4},
 \qquad
 A=2\sqrt Q.
 \label{eq:A-of-Q}
\end{equation}
If $Q$ is uniformly distributed on $[0,1/4]$, the induced density of $A$ is
\begin{equation}
 \rho_A(A)=\rho_Q\!\left(\frac{A^2}{4}\right)
 \frac{\dd}{\dd A}\left(\frac{A^2}{4}\right)
 =2A,
 \qquad0\leq A\leq1.
 \label{eq:partial-measure}
\end{equation}
Thus
\begin{equation}
 \rho_A(A)=2A,\qquad0\leq A\leq1.
 \label{eq:partial-measure-repeat}
\end{equation}
Within the partial-deconfinement ansatz, the Hagedorn ensemble therefore
samples all partially deconfined sectors, with
\begin{equation}
 \Prob\!\left(\frac{M}{N}\in[A,A+\dd A]\right)\simeq2A\,\dd A.
 \label{eq:pd-probability}
\end{equation}
The endpoint $n=N^2/4$ is also the point at which the VKLS profile reaches
the maximal Young-diagram depth allowed by finite rank. This endpoint has been identified in related free matrix models
\cite{BerensteinYan2023}, and the identification of the Young-diagram row
number with $M$ is obtained directly in \cite{Shimada:2026ipv}. Our calculation sharpens that connection by
assigning an explicit canonical probability law to the partially deconfined
fraction.

Concentration of the Young-diagram profile at fixed $M$ is compatible with
the absence of concentration of $M/N$ in the critical canonical ensemble:
the latter averages over the full family with weight $2A\,\dd A$.

The map \eqref{eq:A-of-Q} relies on the stated eigenvalue-density ansatz. The
uniform law for $Q$, by contrast, follows directly from the DG partition
function and does not require the partial-deconfinement interpretation.

\subsection{Tracy--Widom correction at the deconfinement endpoint}
\label{sec:tw-correction}

\paragraph{Tracy--Widom correction to the distribution.}
Recall from \eqref{eq:res1-Z} and \eqref{eq:B0-zero} that
\begin{equation}
 Z_{\DG,N}(1)=\frac{N^2}{4}+\cTW N^{4/3}+o(N^{4/3}),
 \qquad
 \cTW=\frac{|\TWmean|}{2^{4/3}}=B_0(0).
 \nonumber
\end{equation}
In the inner scaling regime, the edge lies at $n=N^2/4$, so its Boltzmann
weight is
\begin{equation}
 \left(1-\frac{t}{N^2}\right)^{N^2/4}\longrightarrow\ee^{-t/4}.
 \label{eq:edge-weight}
\end{equation}
This suggests the refined expansion
\begin{equation}
 Z_{\DG,N}\!\left(1-\frac{t}{N^2}\right)
 =N^2\frac{1-\ee^{-t/4}}{t}
 +\cTW\ee^{-t/4}N^{4/3}+o(N^{4/3}).
 \label{eq:refined-Z}
\end{equation}
Dividing \eqref{eq:refined-Z} by \eqref{eq:res1-Z} gives
\begin{equation}
 \Phihat_N(t)
 =\frac{4(1-\ee^{-t/4})}{t}
 +4\cTW N^{-2/3}
 \left[\ee^{-t/4}-\frac{4(1-\ee^{-t/4})}{t}\right]
 +o(N^{-2/3}).
 \label{eq:laplace-expansion}
\end{equation}
Thus the Tracy--Widom mean controls the leading correction to the full
Laplace transform of the order-parameter distribution. 
Applying the same fixed-edge weighting to \eqref{eq:exact-moment} gives
\begin{equation}
 \langle |P|^{2m}\rangle_{a=1}
 =\frac{1}{(m+1)4^m}
 +\frac{m\cTW}{(m+1)4^{m-1}}N^{-2/3}
 +o(N^{-2/3}),
 \label{eq:res2-moments}
\end{equation}
and in particular
\begin{equation}
 \langle Q\rangle=\frac18+\frac{\cTW}{2}N^{-2/3}+\cdots,
 \qquad
 \langle Q^2\rangle=\frac1{48}+\frac{\cTW}{6}N^{-2/3}+\cdots.
 \label{eq:first-two-moments}
\end{equation}
At the level of the fixed-$t$ expansion \eqref{eq:laplace-expansion}, formally inverting the Laplace transform gives the signed distributional expression
\begin{equation}
 \rho_{Q,N}(q)=4\,\mathbf 1_{[0,1/4]}(q)
 +4\cTW N^{-2/3}
 \left[\delta\!\left(q-\tfrac14\right)-4\,\mathbf 1_{[0,1/4]}(q)\right]
 +\cdots.
 \label{eq:density-correction}
\end{equation}
Equation \eqref{eq:density-correction} should be understood weakly, after integration against suitable test functions, rather than as a pointwise finite-$N$ density. In particular, \eqref{eq:laplace-expansion} by itself controls the exponential test functions $e^{-tq}$; extending the same $N^{-2/3}$ correction to more general test functions requires an additional assumption.
At finite $N$, the delta function in \eqref{eq:density-correction} should be
resolved into a Tracy--Widom boundary layer of width $N^{-2/3}$ in the
variable $q$. Physically, the finite-rank soft edge gives a universal
enhancement near the fully deconfined endpoint.

The endpoint $A=1$ is the fully deconfined end of the partial-deconfinement
 phase. On the
Young-diagram side it is $\lambda'_1=N$, where the VKLS edge reaches the
finite-rank wall. The Tracy--Widom crossover therefore controls the
finite-$N$ weight transferred to the fully deconfined endpoint.
 
The fixed-$t$ expansion determines the integrated endpoint correction but does
not resolve that boundary layer pointwise. Setting $t=sN^{2/3}$ returns the
outer scaling $1-a=s/N^{4/3}$. After the sharp bulk contribution is separated,
Liu's function $B_0(s)$ in \eqref{eq:B0-transform} is the transform of the
centered endpoint profile. The two scaling windows therefore describe the
same wall at different resolutions: fixed $t$ probes the full critical disk,
whereas fixed $s$ resolves its upper edge.

In summary, this section gives the following physical picture:
\begin{enumerate}
\item[(i)] At the Hagedorn point, the normalized Polyakov-loop intensity has
 the large-$N$ limiting law
 \begin{equation}
  \frac{|\tr U|^2}{N^2}\xrightarrow{\ d\ }\Unif\!\left[0,\tfrac14\right].
  \label{eq:summary-i}
 \end{equation}
\item[(ii)] The order parameter is not self-averaging:
 \begin{equation}
  \frac{\Var(|\tr U|^2/N^2)}{\langle|\tr U|^2/N^2\rangle^2}
  \longrightarrow\frac13.
  \label{eq:summary-ii}
 \end{equation}
\item[(iii)] Under the standard partial-deconfinement eigenvalue-density
 ansatz, the deconfined fraction $A=M/N$ has density
 \begin{equation}
  \rho_A(A)=2A,\qquad0\leq A\leq1.
  \label{eq:summary-iii}
 \end{equation}
\item[(iv)] The Tracy--Widom mean governs the first finite-$N$ deformation of
 the entire order-parameter distribution, not only the correction to the
 free energy.
\end{enumerate}
Thus the finite-$N$ smoothing can be read as the edge correction to a broad,
non-self-averaging continuum of partially deconfined sectors.

\section{\texorpdfstring{Finite-$N$ checks}{Finite-N checks}}
\label{sec:numerics}

\subsection{Bessel--Toeplitz representation}
\label{sec:numerical-method}

The finite-rank partition function can be evaluated without sampling unitary
matrices. A Hubbard--Stratonovich transform,
\begin{equation}
 \ee^{a|\tr U|^2}
 =\frac{1}{\pi a}\int\dd^2z\;\ee^{-|z|^2/a}
 \ee^{z\tr U+\bar z\tr U^\dagger},
 \label{eq:HS}
\end{equation}
followed by the Weyl character integral over $U(N)$ gives the
Gross--Witten--Wadia Toeplitz determinant \cite{GrossWitten1980,Wadia1980},
\begin{equation}
 Z_{\DG,N}(a)
 =\frac{2}{a}\int_0^\infty r\,\dd r\,
 \ee^{-r^2/a}D_N(2r),
 \qquad
 D_N(x)=\det[I_{j-k}(x)]_{j,k=1}^N.
 \label{eq:bessel-toeplitz}
\end{equation}
At $a=1$, the equivalent form
\begin{equation}
 Z_{\DG,N}(1)=\frac{N^2}{2}\int_0^\infty g\,\dd g\,
 \ee^{-N^2g^2/4}\det[I_{j-k}(gN)]_{j,k=1}^N
 \label{eq:bessel-toeplitz-g}
\end{equation}
makes the wall region $g\simeq1$ manifest. The normalization is fixed by the
exact value $Z_{\DG,1}(1)=\ee$. 

The same kernel gives the first two factorial moments,
\begin{align}
 \sum_n n\,p_N(n)
 &=2\int_0^\infty r(r^2-1)\ee^{-r^2}D_N(2r)\,\dd r,
 \label{eq:first-kernel}\\
 \sum_n n(n-1)\,p_N(n)
 &=2\int_0^\infty r(r^4-4r^2+2)\ee^{-r^2}D_N(2r)\,\dd r.
 \label{eq:second-kernel}
\end{align}
Thus the free energy, the full fixed-source transform, and the radial moments
are different probes of one and the same finite-$N$ wall measure.

\subsection{Free-energy amplitude}
\label{sec:free-numerics}

\begin{figure}[t]
 \centering
 \includegraphics[width=0.76\linewidth]{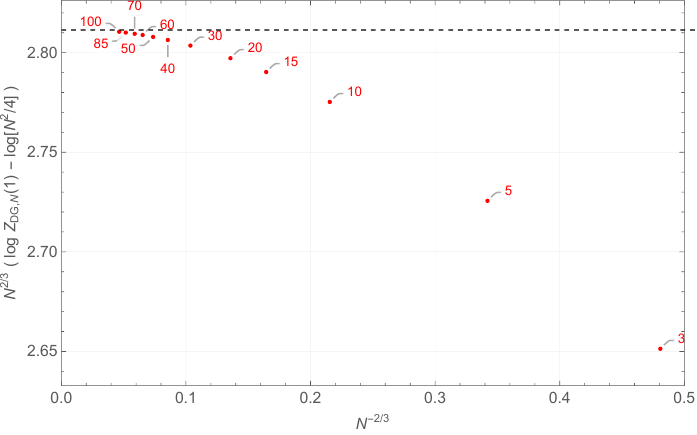}
 \caption{The rescaled critical free-energy residual. The dashed line is the
 Tracy--Widom prediction $4\cTW=2.811425\ldots$.}
 \label{fig:free-energy}
\end{figure}

Define
\begin{equation}
 \Delta_N=\log Z_{\DG,N}(1)-(2\log N-\log4).
 \label{eq:deltaN}
\end{equation}
The soft-edge prediction is
\begin{equation}
 N^{2/3}\Delta_N\longrightarrow4\cTW=2.811425\ldots.
 \label{eq:free-limit}
\end{equation}
Table~\ref{tab:free-energy} and Figure~\ref{fig:free-energy} show that the
finite-rank sequence approaches this value. The result supports the central
identification of the Hagedorn smoothing coefficient with the signed area of
the Plancherel soft edge, rather than with an independent constant of the
matrix integral.

\begin{table}[t]
\centering
\small
\begin{tabular}{rrr@{\qquad}rrr}
\toprule
$N$ & $\log Z_{\DG,N}(1)$ & $N^{2/3}\Delta_N$ &
$N$ & $\log Z_{\DG,N}(1)$ & $N^{2/3}\Delta_N$\\
\midrule
3  & 2.0902 & 2.6610 & 40  & 6.2314 & 2.8064\\
5  & 2.7651 & 2.7266 & 50  & 6.6446 & 2.8079\\
10 & 3.8168 & 2.7752 & 60  & 6.9856 & 2.8088\\
15 & 4.4886 & 2.7903 & 70  & 7.2761 & 2.8095\\
20 & 4.9848 & 2.7972 & 85  & 7.6444 & 2.8101\\
30 & 5.7065 & 2.8035 & 100 & 7.9545 & 2.8105\\
\midrule
\multicolumn{5}{r}{limit $4\cTW=2^{2/3}|\TWmean|$} & 2.8114\\
\bottomrule
\end{tabular}
\caption{The critical free energy and the rescaled residual
$N^{2/3}\Delta_N$.}
\label{tab:free-energy}
\end{table}

\subsection{Critical transform, endpoint crossover, and radial moments}
\label{sec:polyakov-numerics}

\begin{figure}[t]
 \centering
 \includegraphics[width=0.49\linewidth]{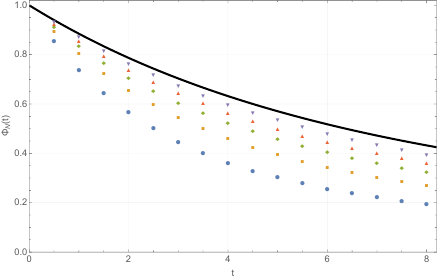}\hfill
 \includegraphics[width=0.49\linewidth]{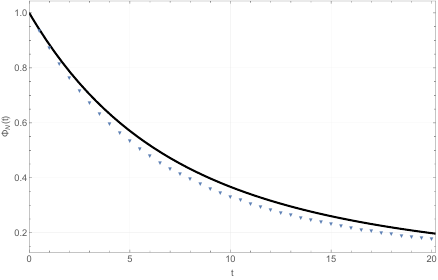}
 \caption{The critical Laplace transform \(\widehat{\Phi}_N(t)=Z_{\mathrm{DG},N}(1-t/N^2)/Z_{\mathrm{DG},N}(1)\). The markers show finite-\(N\) numerical results, and the solid black curve is the uniform-disk prediction \(\widehat{\Phi}_\infty(t)=4(1-e^{-t/4})/t\). Left: \(N=5,10,20,40,100\) over \(0<t\leq8\). Right: \(N=100\) over the extended range \(0<t\leq20\).
 }
 \label{fig:phi}
\end{figure}

\begin{figure}[t]
 \centering
 \includegraphics[width=0.49\linewidth]{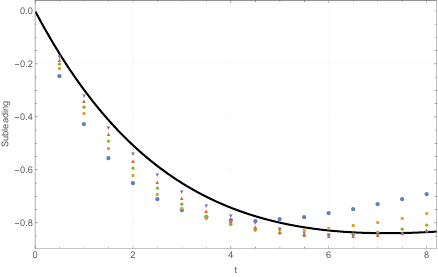}\hfill
 \includegraphics[width=0.49\linewidth]{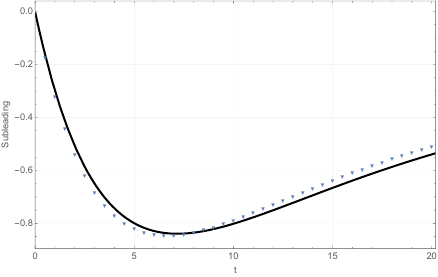}
 \caption{The rescaled correction \(S_N(t)=N^{2/3}[\widehat{\Phi}_N(t)-\widehat{\Phi}_\infty(t)]\) to the critical Laplace transform. The markers show finite-\(N\) numerical results, and the solid black curve is the Tracy–Widom prediction \(S_\infty(t)=4C_{\mathrm{TW}}[e^{-t/4}-\widehat{\Phi}_\infty(t)]\). Left: \(N=5,10,20,40,100\) over \(0<t\leq8\). Right: \(N=100\) over the extended range \(0<t\leq20\).
}
 \label{fig:sub}
\end{figure}

The exact identity \eqref{eq:exact-laplace} tests the complete critical
probability law,
\begin{equation}
 \Phihat_N(t)=\frac{Z_{\DG,N}(1-t/N^2)}{Z_{\DG,N}(1)}
 \longrightarrow\Phihat_\infty(t)=\frac{4(1-\ee^{-t/4})}{t}.
 \label{eq:test-leading}
\end{equation}
The first edge correction is isolated by
\begin{equation}
 S_N(t):=N^{2/3}\left[\Phihat_N(t)-\Phihat_\infty(t)\right]
 \longrightarrow
 S_\infty(t):=4\cTW\left[\ee^{-t/4}-\Phihat_\infty(t)\right].
 \label{eq:test-subleading}
\end{equation}
For fixed $t$, Figure~\ref{fig:phi} and the left panel of
Figure~\ref{fig:sub} show the approach to the uniform-disk transform and to
its Tracy--Widom correction. Since $S_\infty(t)<0$ for $t>0$, the sign has a
direct interpretation: finite rank transfers probability toward the upper
endpoint $Q=1/4$, while the Laplace weight $\ee^{-tQ}$ suppresses that
endpoint more strongly than the interior.

\begin{table}[t]
\centering
\small
\begin{tabular}{r@{\quad}rrrrr@{\quad}r}
\toprule
$t$ & $N=5$ & $N=10$ & $N=20$ & $N=40$ & $N=100$ & $\Phihat_\infty(t)$\\
\midrule
1 & 0.7388 & 0.8004 & 0.8356 & 0.8556 & 0.8698 & 0.8848\\
2 & 0.5648 & 0.6528 & 0.7066 & 0.7386 & 0.7618 & 0.7869\\
4 & 0.3620 & 0.4575 & 0.5228 & 0.5644 & 0.5962 & 0.6321\\
6 & 0.2564 & 0.3407 & 0.4031 & 0.4453 & 0.4787 & 0.5179\\
8 & 0.1956 & 0.2667 & 0.3224 & 0.3615 & 0.3935 & 0.4323\\
\bottomrule
\end{tabular}
\caption{The critical Laplace transform at representative fixed values of
$t$. All entries use the same numerical evaluation and are rounded only in
the displayed table.}
\label{tab:phi}
\end{table}

\begin{table}[t]
\centering
\small
\begin{tabular}{r@{\quad}rrrrr@{\quad}r}
\toprule
$t$ & $N=5$ & $N=10$ & $N=20$ & $N=40$ & $N=100$ & $S_\infty(t)$\\
\midrule
1 & $-0.4270$ & $-0.3916$ & $-0.3625$ & $-0.3410$ & $-0.3224$ & $-0.2980$\\
2 & $-0.6495$ & $-0.6225$ & $-0.5920$ & $-0.5659$ & $-0.5415$ & $-0.5072$\\
4 & $-0.7899$ & $-0.8106$ & $-0.8058$ & $-0.7920$ & $-0.7740$ & $-0.7429$\\
6 & $-0.7646$ & $-0.8226$ & $-0.8456$ & $-0.8499$ & $-0.8454$ & $-0.8288$\\
8 & $-0.6924$ & $-0.7690$ & $-0.8103$ & $-0.8288$ & $-0.8365$ & $-0.8350$\\
\bottomrule
\end{tabular}
\caption{The rescaled residual $S_N(t)$. The mild non-monotonicity at the
largest displayed $t$ is consistent with the crossover from fixed $t$ to
fixed $s=tN^{-2/3}$.}
\label{tab:subleading}
\end{table}

A direct numerical bridge to the outer window is obtained by subtracting the
sharp wall before taking the fixed-$s$ limit. Let
\begin{equation}
 m_N=\left\lceil\frac{N^2}{4}\right\rceil,
 \qquad
 G_N(a)=\sum_{n=0}^{m_N-1}a^n=\frac{1-a^{m_N}}{1-a},
 \label{eq:sharp-geometric}
\end{equation}
and define, by continuity at $s=0$,
\begin{equation}
 \mathcal B_N(s)=\frac{\ee^{sN^{2/3}/4}}{N^{4/3}}
 \left[
 Z_{\DG,N}\!\left(1-\frac{s}{N^{4/3}}\right)
 -G_N\!\left(1-\frac{s}{N^{4/3}}\right)
 \right].
 \label{eq:B0-numeric-extractor}
\end{equation}
The centered wall scaling predicts
$\mathcal B_N(s)\longrightarrow B_0(s)$ for fixed $s$.

The first two radial moments test the same probability law near $t=0$,
\begin{equation}
 \begin{aligned}
 N^{2/3}\left(\langle Q\rangle-\frac18\right)
 &\longrightarrow \frac{\cTW}{2}=0.351428\ldots,\\
 N^{2/3}\left(\langle Q^2\rangle-\frac1{48}\right)
 &\longrightarrow \frac{\cTW}{6}=0.117143\ldots.
 \end{aligned}
 \label{eq:moment-limits}
\end{equation}
Table~\ref{tab:moments} extends both moments to $N=100$. The convergence of the two independent moments to
coefficients fixed by the same $\cTW$ supports the full distributional
statement. In particular,
$\langle Q^2\rangle/\langle Q\rangle^2=1.338$ at $N=100$, approaching the
non-self-averaging limit $4/3$ rather than the self-averaging value one.

\begin{table}[H]
\centering
\small
\begin{tabular}{rrr}
\toprule
$N$ & $N^{2/3}(\langle Q\rangle-1/8)$
    & $N^{2/3}(\langle Q^2\rangle-1/48)$\\
\midrule
5   & 0.5718 & 0.3446\\
10  & 0.4978 & 0.2440\\
20  & 0.4470 & 0.1904\\
40  & 0.4131 & 0.1606\\
100 & 0.3856 & 0.1397\\
\midrule
Tracy--Widom limit & 0.35143 & 0.11714\\
\bottomrule
\end{tabular}
\caption{Rescaled residuals of the first two radial moments. The same
soft-edge moment controls both columns.}
\label{tab:moments}
\end{table}

\subsection{$(a,b)$ model partition function from large-$N$ distribution of the Polyakov loop}
We shall connect the DG model to the phenomenological double-trace deformed $(a,b)$
model of \cite{Alvarez-Gaume:2005dvb}.
The partition function of the $(a,b)$ model is defined as
\begin{align} \label{abmodel}
    Z(a,b) = \int [dU] \exp \left[ a {\rm Tr}U {\rm Tr}U^\dagger +\frac{b}{N^2} \left( {\rm Tr}U {\rm Tr}U^\dagger \right)^2 \right]
\end{align}

To apply the large-$N$ probability distribution of the Polyakov loop at the Hagedorn point (Appendix~\ref{app:largeN-probability}) to the $(a,b)$ model, we consider a narrow window around the Hagedorn phase transition point:
\begin{align}
    a=1-\frac{t}{N^2}, \ \ b= \frac{u}{N^2}
\end{align}
with $t,u \geq 0$. Then, the leading term follows from \eqref{eq:appA-test-function}, while for the first $N^{-2/3}$ correction we additionally assume that the weak expansion \eqref{eq:density-correction} can be tested against $e^{-tq+uq^2}$. Under this assumption, the critical expectation has the expansion
\begin{align}
    \braket{e^{-t Q_N +u Q_N^2}}_{a=1}
    &= \int e^{-tx+ux^2}\rho_Q(x)\,dx \\
    &=4\left(1-4\cTW N^{-2/3}\right)
      \int_0^{1/4}e^{-tx+ux^2}\,dx \nn\\
    &\quad+4\cTW N^{-2/3}e^{-t/4+u/16}+o(N^{-2/3}) \nn\\
    &=4f(t,u)+4\cTW N^{-2/3}
      \left[e^{-t/4+u/16}-4f(t,u)\right]+o(N^{-2/3}),\nn
\end{align}
where we defined
\begin{align}
 f(t,u)&:=\int_0^{1/4}e^{-tx+ux^2}\,dx \nn\\
 &=\frac12e^{-t^2/(4u)}\sqrt{\frac{\pi}{u}}
 \left[\operatorname{Erfi}\!\left(\frac{u-2t}{4\sqrt u}\right)
 +\operatorname{Erfi}\!\left(\frac{t}{2\sqrt u}\right)\right].
\end{align}
The closed form is understood by continuity at $u=0$.

The same expectation value can also be expressed as
\begin{align}
    \braket{e^{-t Q_N +uQ_N^2}}_{a=1}   &=  \frac{1}{Z_{ DG, N}(1)} \int [dU] \exp \left[  \left(  1-\frac{t}{N^2} \right)  |\tr U| ^2 +  \frac{u}{N^4} |\tr U|^4  \right] \\
    &= \frac{ Z \left(  1-\frac{t}{N^2} , \frac{u}{N^2}   \right)  }{Z_{DG,N}(1)}  \nn 
\end{align}
Therefore, the partition function of the $(a,b)$ model can be written as
\begin{align}
    Z  \left( 1- \frac{t}{N^2}, \frac{u}{N^2} \right) &= \braket{e^{-tQ_N + u Q_N^2}}_{a=1} \cdot \frac{N^2}{4} \left( 1+ 4\cTW N^{-2/3} + o(N^{-2/3})    \right)     \\
    &=  \frac{N^2}{4}    \bigg( 4 f(t,u)  + N^{-2/3} 4\cTW e^{-\frac{t}{4} + \frac{u}{16} }    +o( N^{-2/3} )  \bigg)  \nn
\end{align}
The leading term is of order $O(N^2)$ and the subleading term is of order $O(N^{4/3})$.

We first test the leading large $N$ behavior of the partition function. From the expression derived above, we expect
\begin{align}
    \frac{4}{N^2}   Z\left(  1-\frac{t}{N^2},   \frac{u}{N^2}   \right)  = 4f(t,u) + O(N^{-2/3}).
\end{align}

\begin{figure}[t]
    \centering
    \includegraphics[width=\textwidth]{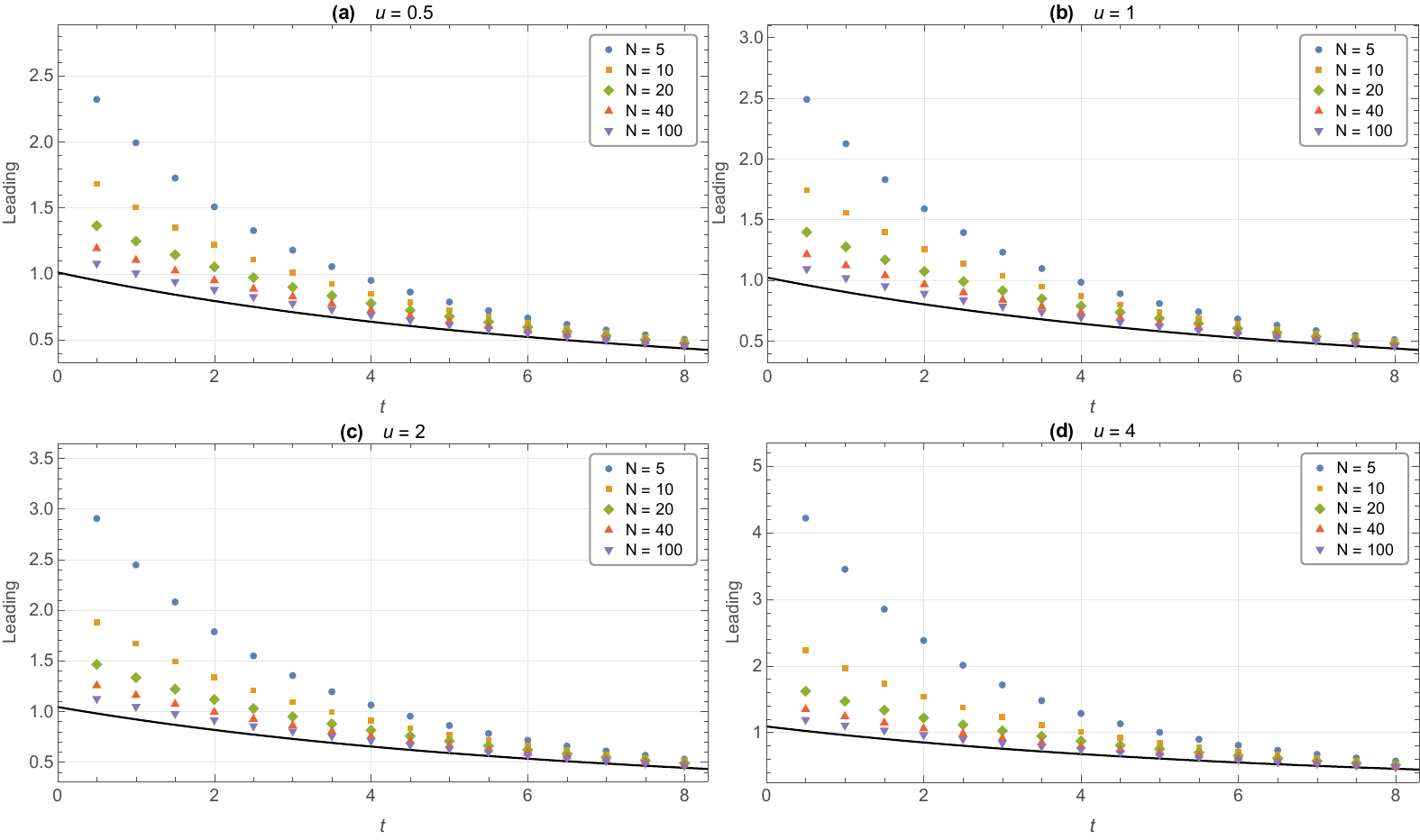}
    \caption{Leading large $N$ behavior of the rescaled partition function. Panels (a), (b), (c), and (d) correspond to $u=0.5$, $u=1$, $u=2$, and $u=4$, respectively. The colored markers show the numerical values of $\frac{4}{N^2} Z\left(1-\frac{t}{N^2},\frac{u}{N^2}\right)$ for $N=5,10,20,40,$ and $100$. The solid black curve is the leading large $N$ prediction $4f(t,u)$.}
    \label{fig:ab-leading}
\end{figure}

Figure~\ref{fig:ab-leading} compares the finite $N$ numerical results with the leading large $N$ prediction $4f(t,u)$ for $u=0.5,1,2,$ and $4$. It shows that the numerical partition function converges to our prediction as $N$ increases.

\begin{figure}[H]
    \centering
    \includegraphics[
        width=\textwidth,
        height=0.78\textheight,
        keepaspectratio
    ]{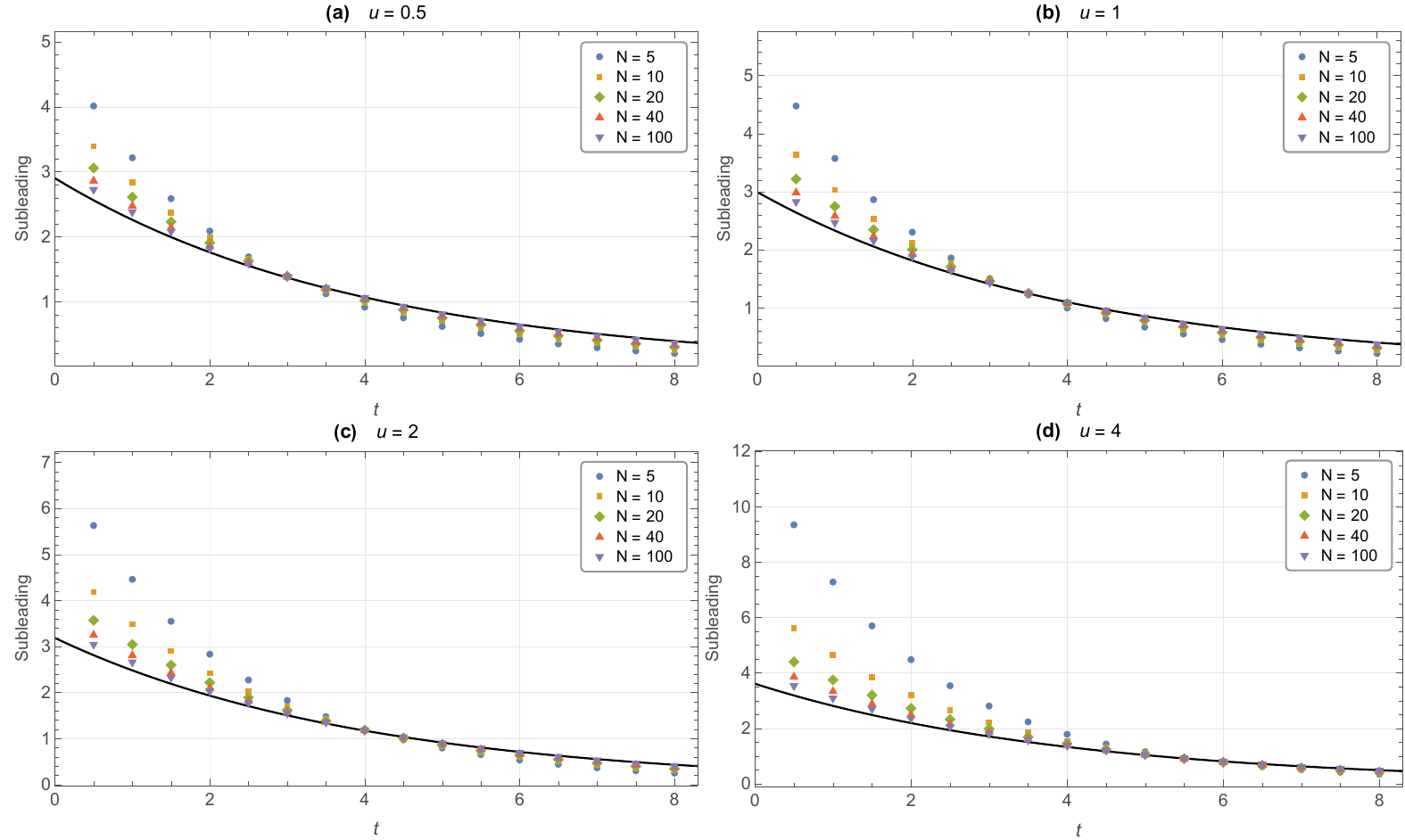}
    \caption{First subleading correction to the rescaled partition function. Panels (a), (b), (c), and (d) correspond to $u=0.5$, $u=1$, $u=2$, and $u=4$, respectively. The colored markers show
    $N^{2/3}\left[ \frac{4}{N^2}
    Z\left(1-\frac{t}{N^2},\frac{u}{N^2}\right)
    -4f(t,u) \right]$ for $N=5,10,20,40,$ and $100$. The solid black curve is the large $N$ prediction $4\cTW e^{-t/4+u/16}$.}
    \label{fig:ab-subleading}
\end{figure}

To examine the first finite $N$ correction, we subtract the leading contribution and rescale the difference by $N^{2/3}$. The quantity
displayed in Fig.~\ref{fig:ab-subleading} is therefore
\begin{align}
    N^{2/3} \left[   \frac{4}{N^2}  Z\left( 1-\frac{t}{N^2},  \frac{u}{N^2} \right) -4f(t,u) \right],
\end{align}
which is predicted to approach
\begin{align}
    4 \cTW e^{-t/4+u/16}
\end{align}
in the large $N$ limit. Figure~\ref{fig:ab-subleading} shows that the numerical subleading correction converges to our prediction as $N$ increases. This agreement provides numerical support for the assumed extension of the weak $N^{-2/3}$ expansion.

\subsection{$(a,b,c)$ model partition function from large-$N$ distribution of the Polyakov loop}

The partition function of the $(a,b,c)$ model is defined as
\begin{align} \label{abcmodel}
    Z(a,b,c) = \int [dU] \exp \left[ a {\rm Tr}U {\rm Tr}U^\dagger +\frac{b}{N^2} \left( {\rm Tr}U {\rm Tr}U^\dagger \right)^2 + \frac{c}{N^4} \left( {\rm Tr}U {\rm Tr}U^\dagger \right)^3 \right]
\end{align}

In the same way as before, we consider a narrow window around the Hagedorn phase transition point:
\begin{align}
    a=1-\frac{t}{N^2}, \ \ b= \frac{u}{N^2}, \ \ c= \frac{v}{N^2}
\end{align}
with $t,u,v \geq 0$. Then, the leading term follows from \eqref{eq:appA-test-function}, while for the first $N^{-2/3}$ correction we additionally assume that the weak expansion \eqref{eq:density-correction} can be tested against $e^{-tq+uq^2+vq^3}$. Under this assumption, the critical expectation has the expansion
\begin{align}
 &\braket{e^{-t Q_N+u Q_N^2+v Q_N^3}}_{a=1}
   =\int e^{-tx+ux^2+vx^3}\rho_Q(x)\,dx \\
 &\quad=4\left(1-4\cTW N^{-2/3}\right)
     \int_0^{1/4}e^{-tx+ux^2+vx^3}\,dx \nn\\
 &\qquad\quad+4\cTW N^{-2/3}e^{-t/4+u/16+v/64}
     +o(N^{-2/3}) \nn\\
 &\quad=4f(t,u,v)+4\cTW N^{-2/3}
     \left[e^{-t/4+u/16+v/64}-4f(t,u,v)\right]
     +o(N^{-2/3}),\nn
\end{align}
where
\begin{align}
 f(t,u,v)&:=\int_0^{1/4}e^{-tx+ux^2+vx^3}\,dx
\end{align}
is a one-dimensional integral over a bounded interval and can be evaluated
numerically.

\begin{figure}[t]
    \centering
    \includegraphics[width=\textwidth]{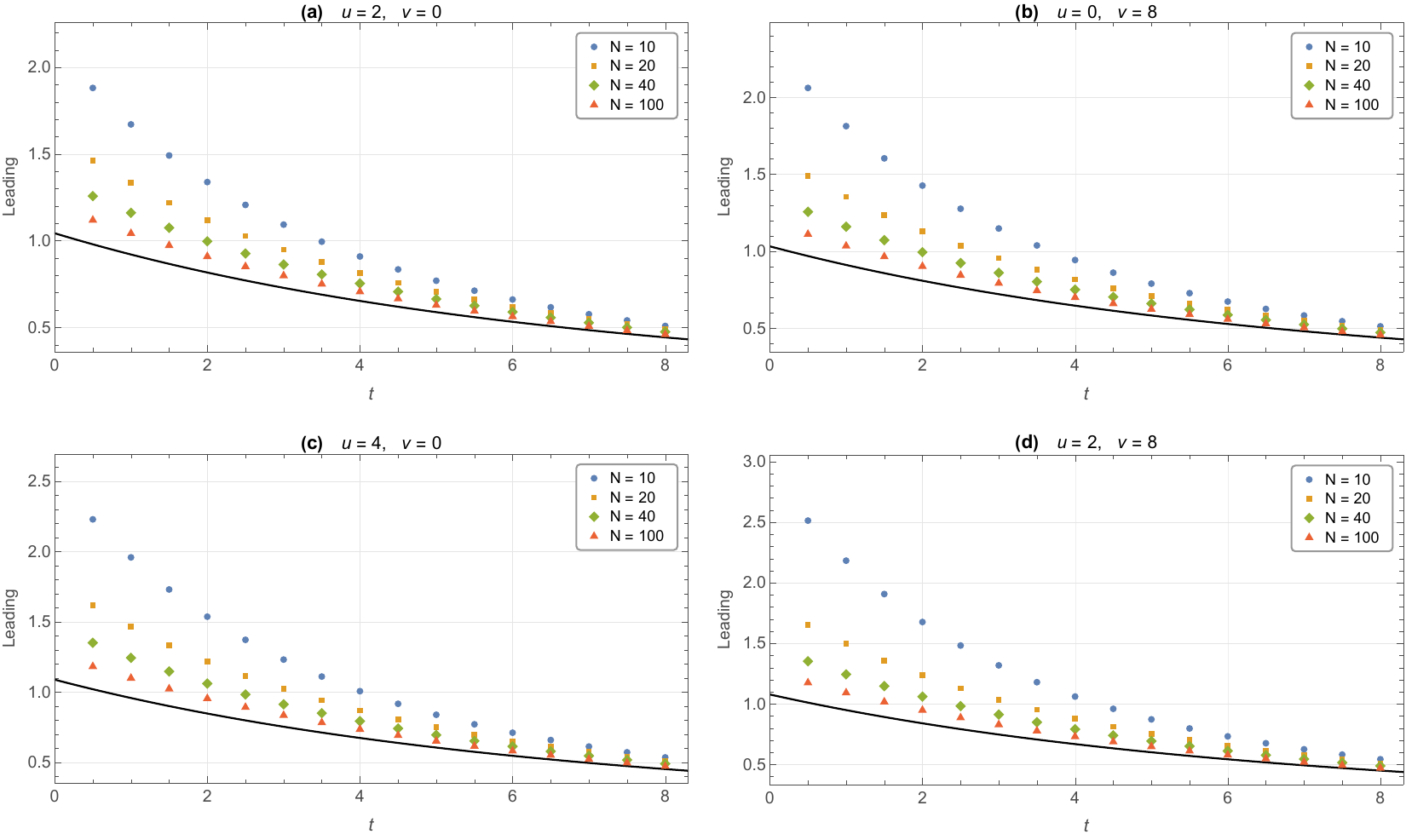}
    \caption{Leading large $N$ behavior of the rescaled partition function of the $(a,b,c)$ model. Panels (a), (b), (c), and (d) correspond to $(u,v)=(2,0)$, $(0,8)$, $(4,0)$, and $(2,8)$, respectively. The colored markers show the numerical values of $\frac{4}{N^2} Z\left(1-\frac{t}{N^2},\frac{u}{N^2},\frac{v}{N^2}\right)$ for $N=10,20,40,$ and $100$. The solid black curve is the leading large $N$ prediction $4f(t,u,v)$, which is different in all four panels.}
    \label{fig:abc-leading}
\end{figure}

\begin{figure}[H]
    \centering
    \includegraphics[
        width=\textwidth,
        height=0.78\textheight,
        keepaspectratio
    ]{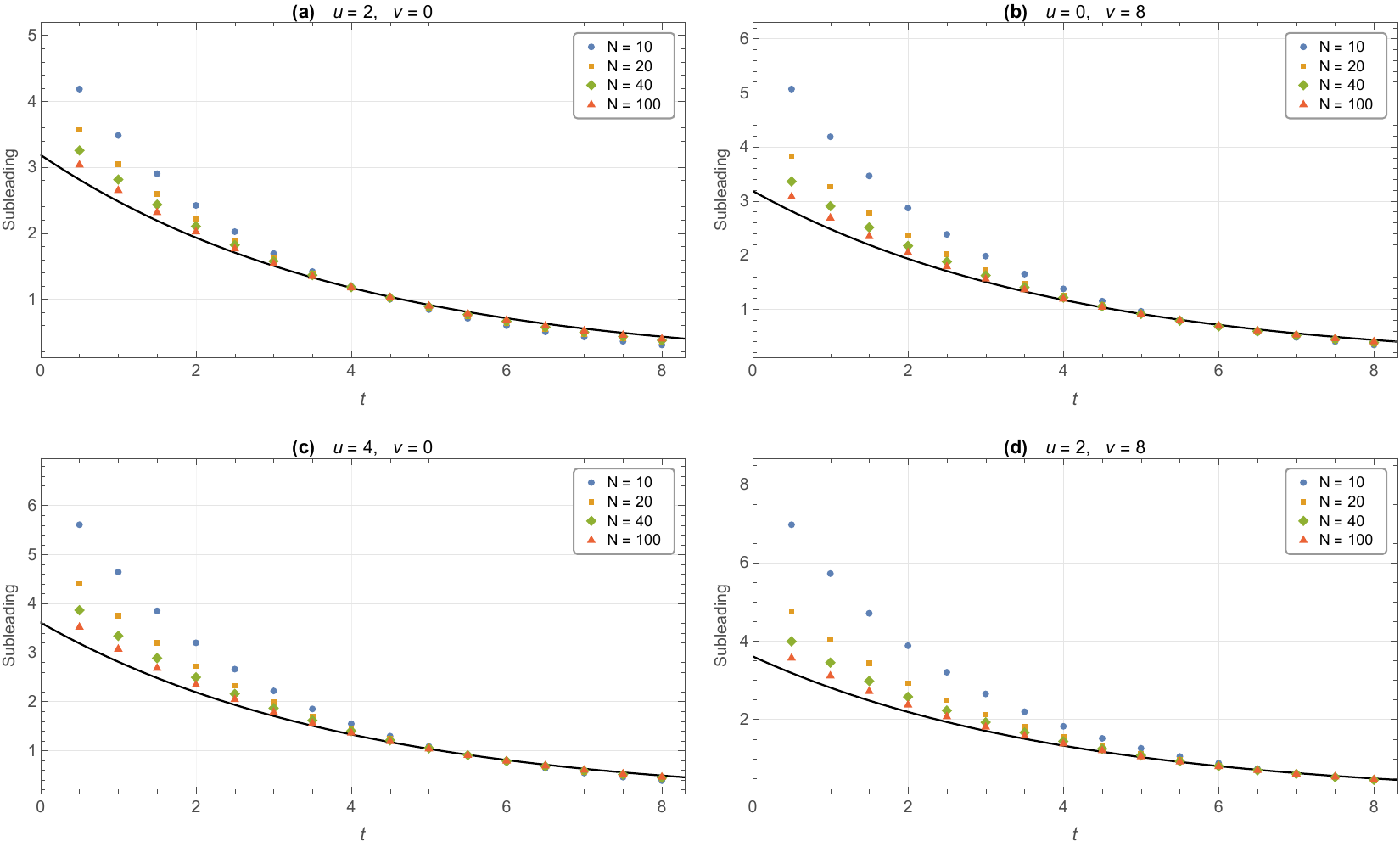}
    \caption{First subleading correction to the rescaled partition function of the $(a,b,c)$ model, for the same four $(u,v)$ as in Fig.~\ref{fig:abc-leading}. The colored markers show
    $N^{2/3}\left[ \frac{4}{N^2}
    Z\left(1-\frac{t}{N^2},\frac{u}{N^2},\frac{v}{N^2}\right)
    -4f(t,u,v) \right]$ for $N=10,20,40,$ and $100$. The solid black curve is the large $N$ prediction $4\cTW e^{-t/4+u/16+v/64}$, which is the same for the two panels in each row.}
    \label{fig:abc-subleading}
\end{figure}

The same expectation value can also be expressed as
\begin{align}
 &\braket{e^{-tQ_N+uQ_N^2+vQ_N^3}}_{a=1}\nn\\
 &\quad=\frac{1}{Z_{\DG,N}(1)}\int[dU]\,
 \exp\!\left[\left(1-\frac{t}{N^2}\right)|\tr U|^2
 +\frac{u}{N^4}|\tr U|^4+\frac{v}{N^6}|\tr U|^6\right]\nn\\
 &\quad=\frac{Z\!\left(1-\frac{t}{N^2},\frac{u}{N^2},\frac{v}{N^2}\right)}
 {Z_{\DG,N}(1)}.
\end{align}
Therefore, the partition function of the $(a,b,c)$ model can be written as
\begin{align}
    Z  \left( 1- \frac{t}{N^2}, \frac{u}{N^2}, \frac{v}{N^2} \right) &= \braket{e^{-tQ_N + u Q_N^2 + v Q_N^3}}_{a=1} \cdot \frac{N^2}{4} \left( 1+ 4\cTW N^{-2/3} + o(N^{-2/3})    \right)     \\
    &=  \frac{N^2}{4}    \bigg( 4 f(t,u,v)  + N^{-2/3} 4 \cTW   e^{-\frac{t}{4} + \frac{u}{16} + \frac{v}{64} }    +o( N^{-2/3} )  \bigg)  \nn
\end{align}
The leading term is of order $O(N^2)$ and the subleading term is of order $O(N^{4/3})$, exactly as in the $(a,b)$ model. The entire $(u,v)$ dependence of the $O(N^{4/3})$ term is the single factor $e^{u/16+v/64}$. Two pairs $(u,v)$ sharing the same $u/16+v/64$ therefore have different leading terms but the same subleading term. We choose the four pairs below so that each row of Fig.~\ref{fig:abc-leading} and Fig.~\ref{fig:abc-subleading} is such a pair.

We first test the leading large $N$ behavior of the partition function. From the expression derived above, we expect
\begin{align}
    \frac{4}{N^2}   Z\left(  1-\frac{t}{N^2},   \frac{u}{N^2},   \frac{v}{N^2}   \right)  = 4f(t,u,v) + O(N^{-2/3}).
\end{align}

Figure~\ref{fig:abc-leading} compares the finite $N$ numerical results with the leading large $N$ prediction $4f(t,u,v)$ for $(u,v)=(2,0)$, $(0,8)$, $(4,0)$, and $(2,8)$. It shows that the numerical partition function converges to our prediction as $N$ increases. We use $N=10,20,40,$ and $100$ here.

To examine the first finite $N$ correction, we subtract the leading contribution and rescale the difference by $N^{2/3}$. The quantity
displayed in Fig.~\ref{fig:abc-subleading} is therefore
\begin{align}
    N^{2/3} \left[   \frac{4}{N^2}  Z\left( 1-\frac{t}{N^2},  \frac{u}{N^2},  \frac{v}{N^2} \right) -4f(t,u,v) \right],
\end{align}
which is predicted to approach
\begin{align}
    4\cTW e^{-t/4+u/16+v/64}
\end{align}
in the large $N$ limit. Figure~\ref{fig:abc-subleading} shows that the numerical subleading correction converges to our prediction as $N$ increases. This agreement again provides numerical support for the assumed extension of the weak $N^{-2/3}$ expansion. Note that panels (a) and (b) share the same prediction $4 \cTW e^{-t/4+1/8}$, and panels (c) and (d) share the same prediction $4 \cTW e^{-t/4+1/4}$, even though their leading terms in Fig.~\ref{fig:abc-leading} are different.

\clearpage

\section{Discussion}
\label{sec:discussion}

The two main results of this paper arise from the same finite-rank
Plancherel wall. In the outer window $1-a\sim N^{-4/3}$, the wall is
resolved on the Tracy--Widom scale, and the signed area of the centered
edge profile gives the leading correction to the critical free energy.
In the inner window $1-a\sim N^{-2}$, the whole interval below the wall
is weighted, producing the uniform critical disk. The edge contribution
then gives the first $N^{-2/3}$ correction to the fixed-source transform
and to every fixed radial moment. The numerical results test both
aspects of this picture.

The same soft-edge moment consequently has two complementary
interpretations. In the outer window, $\cTW$ is the signed area between
the Tracy--Widom edge profile and the sharp wall. In the inner window,
the same coefficient fixes the order-$N^{-2/3}$ probability weight
transferred from the interior of the critical family to its fully
deconfined endpoint. These two interpretations agree because the free
energy and the order-parameter distribution are generated by the same
centered Plancherel-wall sum.

This comparison also separates two levels of the large-$N$ physics.
The leading uniform law follows from the macroscopic VKLS wall and is
therefore a limit-shape effect, whereas the first finite-rank
deformation is controlled by the local BDJ soft edge. Thus the
persistence of the leading uniform measure and the persistence of the
Tracy--Widom endpoint correction are logically distinct questions in
a deformation of the model.

The probability law contains information that is not fixed by the
planar flat potential alone. Planar flatness says that the gapless
family is not separated by an $O(N^2)$ free-energy cost, while the
Schur sum determines the normalized canonical measure on that family.
In partial-deconfinement variables this measure is
\[
  \rho_A(A)=2A,\qquad 0\leq A\leq1.
\]
Equivalently, the measure is uniform in 
$A^2$, rather than in $A$ itself. The canonical
Hagedorn state is therefore a normalized mixture over the
microcanonical partial-deconfinement family, rather than ordinary
coexistence between two isolated saddles. The fully deconfined
endpoint $A=1$ is precisely the Plancherel wall at which the
Tracy--Widom crossover occurs.

The single-winding model also provides a clean baseline for
extensions. The single-winding truncation should be distinguished from a theory with
only one adjoint letter: in the former, the Plancherel weight follows from
the exponential expansion and the Frobenius formula.
For several free adjoint letters, the restriction $\ell(R)\leq N$ remains,
but the representation weights contain multiplicities
\cite{BhattacharyyaCollinsDeMelloKoch2008,BrownHeslopRamgoolam2008,
deMelloKochKimVanZyl2026,deMelloKochRodrigues2026}. At fixed
occupations $n_1,n_2$, for example,
\begin{equation}
 w_R(n_1,n_2)
 =
 \sum_{\substack{R_1\vdash n_1\\R_2\vdash n_2}}
 \bigl(c^R_{R_1R_2}\bigr)^2,
 \qquad |R|=n_1+n_2.
 \label{eq:LR2-weight}
\end{equation}
Here $c^R_{R_1R_2}$ are Littlewood--Richardson coefficients
\cite{BhattacharyyaCollinsDeMelloKoch2008,BerensteinYan2023}.
The microcanonical analysis of \cite{BerensteinYan2023} bears on this question.
That work argues that the two-matrix singlet counting is dominated by VKLS
profiles, despite the squared Littlewood--Richardson weights, and finds
numerical evidence for an endpoint at $n=N^2/4$ independent of the charge
fraction in the regime studied. The persistence of the VKLS profile is also established for the partially
deconfined Gaussian model by Shimada and Watanabe
\cite{Shimada:2026ipv}.
These results support the geometric endpoint, but the row restriction alone
does not determine either the Polyakov-loop support or its normalized
canonical density: the relation between the diagram scale and $|P|$, and
the measure on that scale, must also be specified.
In particular, multiplicities and secondary invariants need not produce a
nonuniform critical density after the occupation numbers have been summed
with their thermal weights. The weakly first-order exponent found in
\cite{BerensteinYan2023} above the endpoint describes the truncated-VKLS regime
and is a different question from the Tracy--Widom smoothing of the endpoint
itself.

For the free Gaussian model with $D>1$ bosonic adjoint
letters, the Hagedorn-point coefficients satisfy $a_1=1$ and
$a_k=D^{1-k}<1$ for $k\geq2$
\cite{Aharony2004,Shimada:2026ipv}.
Thus the higher modes remain stable, and the planar ungapped
saddle family remains flat.
Establishing whether the normalized canonical measure retains
the uniform-disk law requires additional control of higher-mode
fluctuations and finite-rank boundary effects.

The uniform DG law furnishes the base measure for the multi-trace
deformations considered here in their inner double-scaling regime. In the inner double-scaling regime, the $(a,b)$ and $(a,b,c)$ couplings tilt this measure by $e^{-tQ+uQ^2}$ and $e^{-tQ+uQ^2+vQ^3}$, while preserving the finite-rank support $0\le Q\le 1/4$. This picture receives strong numerical support: for both models, the finite-$N$ partition functions converge to the predicted large-$N$ forms, including the universal $N^{-2/3}$ endpoint correction.

Several further extensions follow naturally. A uniform asymptotic
description connecting the inner scale $1-a\sim N^{-2}$ to Liu's
outer scale $1-a\sim N^{-4/3}$ would place the full critical
probability law and its endpoint smoothing in a single framework.
For higher-winding models, the Schur-measure and
free-fermion formulations
\cite{Murthy2023,Shimada:2026ipv}
provide a route to determining the normalized critical measure,
including the effects of the finite-rank boundary.
Finally, a direct finite-$N$
reconstruction of the block fraction $A=M/N$ from the holonomy
eigenvalue density would test the law $\rho_A(A)=2A$ and resolve the
boundary layer near $A=1$ directly in the holonomy ensemble.

\acknowledgments

We thank Seok Kim for valuable inputs.
This work is supported by a start-up research fund of Huzhou Normal
University (RdMK, MK), a Zhejiang Province talent award and a Changjiang Scholar award
(RdMK), and by the NRF grants funded by the Korea government (MSIT)
(RS-2025-25414114 (MK) and RS-2026-25477643 (HO)).

\appendix

\section{\texorpdfstring{Large-$N$ probability distribution}{Large-N probability distribution} of the Polyakov loop}
\label{app:largeN-probability}

By the large-$N$ probability distribution of the Polyakov loop, we mean the
limiting distribution obtained when $U$ is itself sampled from the canonical
matrix-model measure. At the Hagedorn point $a=1$, the DG model defines the
probability measure
\begin{equation}
 \dd\mu_N(U)=\frac{1}{Z_{\DG,N}(1)}[\dd U]\,\ee^{|\tr U|^2}.
 \label{eq:appA-measure}
\end{equation}
So $U$ is a random unitary matrix, distributed according to $\dd\mu_N$.
Consequently, $P_N=\tr U/N$ is a random variable --- a complex-valued random
variable. For each finite $N$, therefore, there is some probability
distribution for $P_N$. The large-$N$ distribution asks whether these
finite-$N$ distributions approach a limiting probability law as $N\to\infty$.
More precisely, if $\nu_N$ denotes the probability measure induced on the
complex $P$-plane by
\begin{equation}
 U\longmapsto P_N=\frac{\tr U}{N},
 \label{eq:appA-map}
\end{equation}
then we are asking whether $\nu_N$ converges weakly to a limiting
probability measure $\nu_\infty$. Equivalently, for a
suitable test function $f$, we are asking if
\begin{equation}
 \left\langle f\!\left(\frac{\tr U}{N}\right)\right\rangle_{a=1}
 \longrightarrow\int_{\mathbb C}f(P)\,\dd\nu_\infty(P).
 \label{eq:appA-test-function}
\end{equation}
This is much more than computing $\langle P\rangle$, which tells you only one
moment of the distribution. Indeed, in this model there is a $U(1)$ symmetry
$U\to\ee^{i\alpha}U$, under which $\tr U\to\ee^{i\alpha}\tr U$, while the
action depends only on $|\tr U|^2$. Therefore,
\begin{equation}
 \langle P_N\rangle=0
 \label{eq:appA-mean-zero}
\end{equation}
for every $N$. This does not imply that $P_N$ itself is small. Its phase can
average to zero while its magnitude remains $O(1)$. That is why the
distribution is interesting.

We were considering
\begin{equation}
 Q_N=|P_N|^2=\frac{|\tr U|^2}{N^2}.
 \label{eq:appA-QN}
\end{equation}
The claim is that at $a=1$,
\begin{equation}
 Q_N\xrightarrow{\ d\ }Q,
 \qquad
 Q\sim\Unif\!\left[0,\tfrac14\right].
 \label{eq:appA-uniform-Q}
\end{equation}
Thus, for large $N$, if we repeatedly sample $U$ from the canonical ensemble
and compute
\begin{equation}
 q=\frac{|\tr U|^2}{N^2},
 \label{eq:appA-q}
\end{equation}
then $q$ does not pile up at one particular value. Instead its probability
density tends to
\begin{equation}
 \rho_Q(q)=4,\qquad0\leq q\leq\frac14.
 \label{eq:appA-density}
\end{equation}
So every interval of equal width in $q$ has equal probability. For example,
\begin{equation}
 \Prob\!\left(0<Q<\tfrac18\right)\longrightarrow\frac12,
 \qquad\text{and}\qquad
 \Prob\!\left(\tfrac18<Q<\tfrac14\right)\longrightarrow\frac12.
 \label{eq:appA-halves}
\end{equation}
This is very different from the more usual large-$N$ situation in which an
order parameter becomes sharply concentrated around the value determined by
a saddle.

Writing $R=|P|$, we have $Q=R^2$. Since $Q$ is uniform on $[0,1/4]$,
\begin{equation}
 \rho_R(r)=\rho_Q(r^2)\,\frac{\dd(r^2)}{\dd r}=8r,
 \qquad0\leq r\leq\frac12.
 \label{eq:appA-radial}
\end{equation}
Thus
\begin{equation}
 \rho_{|P|}(r)=8r,\qquad0\leq r\leq\frac12.
 \label{eq:appA-radial-repeat}
\end{equation}
In particular,
\begin{equation}
 \langle|P|\rangle=\int_0^{1/2}8r^2\,\dd r=\frac13,
 \qquad\text{while}\qquad
 \langle|P|^2\rangle=\frac18.
 \label{eq:appA-absP}
\end{equation}
So although $\langle P\rangle=0$, a typical configuration has a Polyakov loop
of order one. In fact, there is an especially nice statement for the full
complex Polyakov loop. Because the action is invariant under
$P\to\ee^{i\alpha}P$, the phase of $P$ is uniformly distributed. Combine this
with $\rho_R(r)=8r$. If $P=r\ee^{i\phi}$, then ($\dd\Prob$ means an
infinitesimal probability)
\begin{equation}
 \dd\Prob=8r\,\dd r\,\frac{\dd\phi}{2\pi}
 =\frac4\pi\,r\,\dd r\,\dd\phi.
 \label{eq:appA-plane-measure}
\end{equation}
But $r\,\dd r\,\dd\phi$ is just the measure on a plane so that the limiting
two-dimensional density is simply
\begin{equation}
 \rho_\infty(P)=\frac4\pi\,\mathbf 1_{\{|P|\leq1/2\}}.
 \label{eq:appA-disk-density}
\end{equation}
So a stronger formulation of our result is that $\tr U/N$ becomes
uniformly distributed over the disk $|P|\leq\tfrac12$ at the Hagedorn point.
That is more than saying that $|P|^2$ is uniform.

This is unusual at large $N$. Ordinarily, large $N$ suppresses fluctuations.
One expects an observable $P_N$ to approach a definite saddle-point value,
$P_N\to P_*$, so that its probability distribution becomes
\begin{equation}
 \rho_N(P)\longrightarrow\delta(P-P_*).
 \label{eq:appA-delta}
\end{equation}
Our result says exactly the opposite happens at the Hagedorn point:
\begin{equation}
 \rho_N(P)\longrightarrow\frac4\pi\,\mathbf 1_{\{|P|\leq1/2\}}.
 \label{eq:appA-opposite}
\end{equation}
The distribution retains an $O(1)$ width even at $N=\infty$. There is
therefore no large-$N$ self-averaging of the Polyakov loop at criticality. At
the Hagedorn point, rather than selecting one macroscopic Polyakov-loop
saddle, the canonical ensemble samples a continuum of macroscopic
Polyakov-loop values, uniformly over a disk in the complex $P$-plane.

If the identification $|P|=A/2$, with $A=M/N$ the partially deconfined
fraction, is justified for these configurations, this continuum becomes a
continuum of partially deconfined sectors.

\section{Uniform large-deviation control away from the Tracy--Widom edge}
\label{app:uniform-large-deviation}

Our derivation of the uniform law
\begin{equation}
 Q_N=\frac{|\tr U|^2}{N^2}\xrightarrow{\ d\ }\Unif\!\left[0,\tfrac14\right]
 \label{eq:appB-uniform}
\end{equation}
uses an approximation which is very plausible, but stronger than what the
Tracy--Widom theorem by itself tells us. The key step is to replace
$p_N(n)=\Prob_n(\lambda'_1\leq N)$ by a sharp step function,
\begin{equation}
 p_N(n)\approx
 \begin{cases}
 1,&n<N^2/4,\\[2pt]
 0,&n>N^2/4.
 \end{cases}
 \label{eq:appB-step}
\end{equation}
We used this to write
\begin{equation}
 Z_{\DG,N}\!\left(1-\frac{t}{N^2}\right)
 =\sum_{n\geq0}\left(1-\frac{t}{N^2}\right)^n p_N(n)
 \label{eq:appB-exact-sum}
\end{equation}
and approximated it by
\begin{equation}
 \sum_{n=0}^{N^2/4}\left(1-\frac{t}{N^2}\right)^n.
 \label{eq:appB-approx-sum}
\end{equation}
To prove this rigorously, we need to consider the error made in this
replacement and show it vanishes.

What does Tracy--Widom tell us? Near
\begin{equation}
 n_*=\frac{N^2}{4},
 \label{eq:appB-nstar}
\end{equation}
we can write
\begin{equation}
 n=n_*+N^{4/3}\tau.
 \label{eq:appB-tau}
\end{equation}
Then BDJ gives
\begin{equation}
 p_N(n)\longrightarrow F_2(-2^{4/3}\tau)
 \label{eq:appB-bdj}
\end{equation}
for fixed $\tau$. So it gives extremely precise information in the window
\begin{equation}
 |n-n_*|=O(N^{4/3}).
 \label{eq:appB-window}
\end{equation}
But our uniform-law argument sums over all $n=0,1,2,\ldots$, including values
separated from $n_*$ by distances of order $N^2$. That is the issue.

It is easy to see why pointwise statements are not enough. Suppose we know
that for every fixed macroscopic $x<1/4$,
\begin{equation}
 p_N(N^2x)\longrightarrow1.
 \label{eq:appB-below-pointwise}
\end{equation}
And for every fixed $x>1/4$,
\begin{equation}
 p_N(N^2x)\longrightarrow0.
 \label{eq:appB-above-pointwise}
\end{equation}
This already strongly suggests
\begin{equation}
 p_N(N^2x)\longrightarrow\mathbf 1_{[0,1/4]}(x).
 \label{eq:appB-indicator}
\end{equation}
Pointwise convergence alone does not control the infinite range of the
sum.
The relevant requirement is that the total error be $o(N^2)$; for example,
an error of $1/N$ in each of $O(N^2)$ terms is already negligible after
normalization by $N^2$.
Uniform control on macroscopic intervals and a summable far-tail bound are
sufficient, as shown in Appendix~\ref{app:weak-concentration}.

We can achieve this with a large-deviation estimate which can give bounds of
roughly the following form: Take a region safely below the wall, say
\begin{equation}
 n\leq\frac{N^2}{4}-CN^{4/3}
 \label{eq:appB-below-edge}
\end{equation}
or, even at macroscopic separation,
\begin{equation}
 n\leq\left(\frac14-\epsilon\right)N^2.
 \label{eq:appB-below-macro}
\end{equation}
There we want to know that
\begin{equation}
 1-p_N(n)=\Prob_n(\lambda'_1>N)
 \label{eq:appB-below}
\end{equation}
is not merely tending to zero, but is uniformly very small throughout this
region. Likewise, above the wall,
\begin{equation}
 n\geq\left(\frac14+\epsilon\right)N^2,
 \label{eq:appB-above-macro}
\end{equation}
we want
\begin{equation}
 p_N(n)=\Prob_n(\lambda'_1\leq N)
 \label{eq:appB-above}
\end{equation}
to be uniformly very small. Ideally one has large-deviation estimates
schematically like
\begin{equation}
 1-p_N(n)\lesssim\ee^{-N^\alpha I_-(n/N^2)}
 \label{eq:appB-ldp-below}
\end{equation}
below the edge, and
\begin{equation}
 p_N(n)\lesssim\ee^{-N^\alpha I_+(n/N^2)}
 \label{eq:appB-ldp-above}
\end{equation}
above the edge, for some positive rate functions $I_\pm$. The exact exponent
and rate function aren't important --- we just need that these errors decay
sufficiently rapidly that summing over $O(N^2)$ possible $n$'s still produces
a negligible error.

A clean proof would divide
\begin{equation}
 Z_{\DG,N}\!\left(1-\frac{t}{N^2}\right)
 \label{eq:appB-ZN}
\end{equation}
into three regions: below the edge, at the Tracy--Widom edge and above the
edge. For example, choose $L=L_N\to\infty$ with $L_N=o(N^{2/3})$ and write
\begin{equation}
 n_\pm=\frac{N^2}{4}\pm LN^{4/3}.
 \label{eq:appB-npm}
\end{equation}
Then break the sum up as follows
\begin{equation}
 \sum_{n<n_-}+\sum_{n_-\leq n\leq n_+}+\sum_{n>n_+}.
 \label{eq:appB-three-regions}
\end{equation}
In the first region we want $p_N(n)\simeq1$ uniformly. In the third we want
$p_N(n)\simeq0$ uniformly. The middle region is the Tracy--Widom region.

The middle region contains $O(L_NN^{4/3})$ terms. Since every term satisfies $0\leq p_N(n)\leq1$, its total contribution is
at most $O(L_NN^{4/3})$. But the leading partition function is
\begin{equation}
 Z_{\DG,N}(1)\sim\frac{N^2}{4}.
 \label{eq:appB-Z1}
\end{equation}
Therefore, after dividing by $N^2$, the whole edge region is bounded by a constant times
\begin{equation}
 \frac{L_NN^{4/3}}{N^2}=L_NN^{-2/3}\longrightarrow0.
 \label{eq:appB-edge-fraction}
\end{equation}
This is why the precise Tracy--Widom shape is irrelevant to the leading
uniform law. It becomes relevant only for the first $N^{-2/3}$ correction.

The leading result would follow if we can establish
\begin{equation}
 \frac1{N^2}\sum_{n\geq0}
 \left|p_N(n)-\mathbf 1_{\{n<N^2/4\}}\right|\longrightarrow0.
 \label{eq:appB-L1-target}
\end{equation}
This equation captures the issue very neatly. If this holds, then
\begin{equation}
 \frac1{N^2}Z_{\DG,N}\!\left(1-\frac{t}{N^2}\right)
 =\frac1{N^2}\sum_n\left(1-\frac{t}{N^2}\right)^n p_N(n)
 \longrightarrow\int_0^{1/4}\ee^{-tx}\dd x.
 \label{eq:appB-consequence}
\end{equation}
This result then gives
\begin{equation}
 \frac{Z_{\DG,N}(1-t/N^2)}{Z_{\DG,N}(1)}\longrightarrow4\int_0^{1/4}\ee^{-tx}\dd x,
 \label{eq:appB-ratio}
\end{equation}
which is the Laplace transform of the uniform distribution.

BDJ tells us beautifully what happens inside the narrow crossover region, but
to turn the intuitive step-function picture into a theorem for the complete
$N^2$-scale sum, one also needs quantitative bounds guaranteeing that
$p_N(n)$ is uniformly close to $1$ below the crossover and uniformly close to
$0$ above it.

\section{A weaker concentration estimate}
\label{app:weak-concentration}

Let's start from VKLS. Recall $p_N(n)=\Prob_n(\lambda'_1\leq N)$. VKLS says
\begin{equation}
 \frac{\lambda'_1}{\sqrt n}\xrightarrow{\ \Prob\ }2.
 \label{eq:appC-vkls}
\end{equation}
The wall reaches the edge at $n_*=N^2/4$. This macroscopic concentration is standard Plancherel limit-shape behavior
\cite{VershikKerov1977,LoganShepp1977,BorodinOkounkovOlshanski2000};
BDJ refines it on the $n^{1/6}$ edge scale and proves Tracy--Widom
convergence, including convergence of moments \cite{BaikDeiftJohansson1999}. Fix any $\epsilon>0$, and define
\begin{equation}
 n_-=\left(\frac14-\epsilon\right)N^2,
 \qquad
 n_+=\left(\frac14+\epsilon\right)N^2.
 \label{eq:appC-npm}
\end{equation}
At $n_-$,
\begin{equation}
 \frac{N}{\sqrt{n_-}}\longrightarrow\frac{1}{\sqrt{1/4-\epsilon}}>2.
 \label{eq:appC-below-ratio}
\end{equation}
Since $\lambda'_1/\sqrt n\to2$ in probability,
\begin{equation}
 p_N(n_-)=\Prob\!\left(\frac{\lambda'_1}{\sqrt{n_-}}
 \leq\frac{N}{\sqrt{n_-}}\right)\longrightarrow1.
 \label{eq:appC-below-limit}
\end{equation}
Similarly,
\begin{equation}
 \frac{N}{\sqrt{n_+}}\longrightarrow\frac{1}{\sqrt{1/4+\epsilon}}<2,
 \label{eq:appC-above-ratio}
\end{equation}
so $p_N(n_+)\to0$. Thus we only need the ordinary VKLS law of large numbers
\begin{equation}
 p_N\!\left((1/4-\epsilon)N^2\right)\to1,
 \qquad
 p_N\!\left((1/4+\epsilon)N^2\right)\to0.
 \label{eq:appC-two-points}
\end{equation}
No rate is needed.

Now we will argue that monotonicity makes the convergence uniform away from
the wall. Through Plancherel growth,
\begin{equation}
 p_N(n)=\Prob(T_N>n),
 \label{eq:appC-hitting-time}
\end{equation}
where $T_N$ is the first time the diagram acquires $N+1$ rows. Therefore
$p_N(n)$ is non-increasing in $n$. Consequently, for every $n\leq n_-$,
\begin{equation}
 1-p_N(n)\leq1-p_N(n_-)=o(1).
 \label{eq:appC-below-pointwise}
\end{equation}
Hence\footnote{Recall that $o(N^2)$ means ``smaller than $N^2$ by a factor
that goes to zero as $N\to\infty$.'' On the other hand $O(N^2)$ means
``bounded by a constant times $N^2$ for sufficiently large $N$.''}
\begin{equation}
 \sum_{n\leq n_-}[1-p_N(n)]\leq n_-[1-p_N(n_-)]=o(N^2).
 \label{eq:appC-below-sum}
\end{equation}
This gives uniform control over the entire region below the wall. Likewise,
for $n\geq n_+$,
\begin{equation}
 p_N(n)\leq p_N(n_+)=o(1).
 \label{eq:appC-above-pointwise}
\end{equation}
Therefore, on any interval $n_+\leq n\leq CN^2$, we have
\begin{equation}
 \sum_{n=n_+}^{CN^2}p_N(n)\leq CN^2p_N(n_+)=o(N^2).
 \label{eq:appC-above-sum}
\end{equation}
So monotonicity converts two pointwise concentration statements into uniform
control over $O(N^2)$ terms.

We now want to argue that the small region around the wall costs only
$O(\epsilon N^2)$. Between $n_-$ and $n_+$ there are only
\begin{equation}
 n_+-n_-=2\epsilon N^2+O(1)
 \label{eq:appC-middle-count}
\end{equation}
terms. Since $0\leq p_N(n)\leq1$, we do not need to know anything at all
about the detailed behavior in this region
\begin{equation}
 \sum_{n_-<n<n_+}
 \left|p_N(n)-\mathbf 1_{\{n<N^2/4\}}\right|
 \leq2\epsilon N^2+O(1).
 \label{eq:appC-middle}
\end{equation}
Divide by $N^2$ to find
\begin{equation}
 \frac1{N^2}\,(\text{middle contribution})\leq2\epsilon+o(1).
 \label{eq:appC-middle-fraction}
\end{equation}
We now take $\epsilon\to0$ so the middle terms contribute nothing.

One place where we should be careful is with the infinite tail $n\gg N^2$.
The bound \eqref{eq:absolute-bound} is sufficient
\begin{equation}
 p_N(n)\leq\frac{N^{2n}}{n!}.
 \label{eq:appC-tail-bound}
\end{equation}
This follows from the Haar-integral representation and $|\tr U|\leq N$. Take
a fixed constant $C>\ee$, and put $m=\lceil CN^2\rceil$. Using
\begin{equation}
 m!\geq\left(\frac{m}{\ee}\right)^m,
 \label{eq:appC-stirling}
\end{equation}
we find
\begin{equation}
 \frac{N^{2m}}{m!}\leq\left(\frac{\ee N^2}{m}\right)^m
 \lesssim\left(\frac{\ee}{C}\right)^{CN^2}.
 \label{eq:appC-tail-estimate}
\end{equation}
Since $C>\ee$, we have $\ee/C<1$, so this is exponentially small in $N^2$.
Moreover, for $n\geq m$,
\begin{equation}
 \frac{N^{2(n+1)}/(n+1)!}{N^{2n}/n!}=\frac{N^2}{n+1}\leq\frac1C+o(1)<1.
 \label{eq:appC-ratio}
\end{equation}
Thus the whole tail is bounded by a geometric series:
\begin{equation}
 \sum_{n\geq CN^2}p_N(n)
 \leq O\!\left[\left(\frac{\ee}{C}\right)^{CN^2}\right]=o(N^2).
 \label{eq:appC-far-tail}
\end{equation}
So even the far tail requires no sophisticated Plancherel large-deviation
theorem.

Now we just need to put everything together. Define
\begin{equation}
 D_N=\sum_{n\geq0}\left|p_N(n)-\mathbf 1_{\{n<N^2/4\}}\right|.
 \label{eq:appC-DN}
\end{equation}
Split the sum into
\begin{equation}
 [0,n_-],\quad[n_-,n_+],\quad[n_+,CN^2],\quad[CN^2,\infty).
 \label{eq:appC-regions}
\end{equation}
The four estimates above give
\begin{equation}
 D_N\leq o(N^2)+2\epsilon N^2+o(N^2)+o(N^2).
 \label{eq:appC-assembly}
\end{equation}
Therefore
\begin{equation}
 \limsup_{N\to\infty}\frac{D_N}{N^2}\leq2\epsilon.
 \label{eq:appC-limsup}
\end{equation}
Since $\epsilon>0$ was arbitrary, we have the result we wanted
\begin{equation}
 \frac1{N^2}\sum_{n\geq0}
 \left|p_N(n)-\mathbf 1_{\{n<N^2/4\}}\right|\longrightarrow0.
 \label{eq:appC-L1}
\end{equation}
This immediately proves the inner scaling result. Take
\begin{equation}
 a_N=1-\frac{t}{N^2},\qquad t\geq0
 \label{eq:appC-aN}
\end{equation}
with $t$ fixed. Since $0\leq a_N^n\leq1$,
\begin{equation}
 \left|\frac1{N^2}\sum_{n\geq0}a_N^n\,p_N(n)
 -\frac1{N^2}\sum_{n<N^2/4}a_N^n\right|
 \leq\frac1{N^2}\sum_{n\geq0}
 \left|p_N(n)-\mathbf 1_{\{n<N^2/4\}}\right|
 \longrightarrow0.
 \label{eq:appC-weighted-L1}
\end{equation}
Using
\begin{equation}
 \frac1{N^2}\sum_{n<N^2/4}\left(1-\frac{t}{N^2}\right)^n
 \longrightarrow\int_0^{1/4}\ee^{-tx}\dd x
 \label{eq:appC-riemann}
\end{equation}
we have
\begin{equation}
 \frac1{N^2}Z_{\DG,N}\!\left(1-\frac{t}{N^2}\right)
 \longrightarrow\frac{1-\ee^{-t/4}}{t}.
 \label{eq:appC-inner-limit}
\end{equation}
At $t=0$, this gives
\begin{equation}
 \frac{Z_{\DG,N}(1)}{N^2}\longrightarrow\frac14.
 \label{eq:appC-Z1}
\end{equation}
So we do not even need to assume the critical normalization separately.
Taking the ratio,
\begin{equation}
 \frac{Z_{\DG,N}(1-t/N^2)}{Z_{\DG,N}(1)}
 \longrightarrow\frac{4(1-\ee^{-t/4})}{t}.
 \label{eq:appC-ratio-limit}
\end{equation}
Finally, since
\begin{equation}
 \frac{Z_{\DG,N}(1-t/N^2)}{Z_{\DG,N}(1)}
 =\left\langle\ee^{-t|\tr U|^2/N^2}\right\rangle_{a=1},
 \label{eq:appC-laplace-identity}
\end{equation}
this is the Laplace transform of $Q\sim\Unif[0,1/4]$. Hence
\begin{equation}
 \frac{|\tr U|^2}{N^2}\xrightarrow{\ d\ }\Unif\!\left[0,\tfrac14\right].
 \label{eq:appC-final}
\end{equation}
The leading uniform law requires only the VKLS concentration in probability
together with monotonicity of the Plancherel wall probability and a crude
bound on the far tail. No uniform large-deviation estimate is needed. The
detailed Tracy--Widom $N^{4/3}$ edge structure enters only in computing the
first $N^{-2/3}$ correction to this limiting law. That last distinction is
important: the uniform disk law is a limit-shape phenomenon; Tracy--Widom
controls how finite $N$ approaches that law.

\section{Conventions and normalization}
\label{app:conventions}

Let $X$ have distribution function $F$ and a finite first moment. Integration
by parts on the two half-lines gives
\begin{equation}
 \int_{-\infty}^{\infty}[F(x)-\theta(x)]\dd x=-\langle X\rangle.
 \label{eq:cdf-moment}
\end{equation}
This is the identity used in \eqref{eq:tw-edge-integral}.

For the Painlev\'e-II normalization, let $u(x)$ be the Hastings--McLeod solution
\begin{equation}
 u''=2u^3+xu,
 \qquad
 u(x)\sim\operatorname{Ai}(x)\quad(x\longrightarrow+\infty).
\end{equation}
The GUE Tracy--Widom CDF satisfies
\begin{equation}
 \frac{\dd^2}{\dd x^2}\log F_2(x)=-u(x)^2,
 \qquad
 F_2(x)=\exp\!\left[-\int_x^\infty(y-x)u(y)^2\dd y\right]
 \label{eq:TW-Painleve}
\end{equation}
\cite{TracyWidom1994,HastingsMcLeod1980}. Liu's convention
$\tfrac12f''=f^3+tf$ is obtained with
$f(t)=2^{1/3}u(2^{1/3}t)$. Together with the common boundary condition at
$+\infty$, this gives
\begin{equation}
 \ee^{F^{(2)}_0(t)}=F_2(2^{1/3}t),
\end{equation}
which fixes the powers of two in \eqref{eq:B0-transform}.

\end{document}